\documentclass[fp,twocolumn]{jpsj3}
\usepackage{txfonts}
\usepackage{color}

\title{Optimal Thermoelectric Power Factor of Narrow-Gap Semiconducting Carbon Nanotubes with Randomly Substituted Impurities}

\author{Manaho Matsubara$^{1}$, Kenji Sasaoka$^{2}$, Takahiro Yamamoto$^{1,2}$, and Hidetoshi Fukuyama$^{3}$}
\inst{$^{1}$Department of Physics, Faculty of Science, Tokyo University of Science, Shinjuku, Tokyo 162-8601, Japan\\
$^{2}$W-FST Research Center, RIST, Tokyo University of Science, Shinjuku, Tokyo 162-8601, Japan \\
$^{3}$Tokyo University of Science, Shinjuku, Tokyo 162-8601, Japan} 

\abst{We have theoretically investigated thermoelectric (TE) effects of narrow-gap single-walled carbon nanotubes (SWCNTs) with randomly substituted nitrogen (N) impurities, i.e., N-substituted (20,0) SWCNTs with a band gap of 0.497 eV.
For such a narrow-gap system, the thermal excitation from the valence band to the conduction band contributes to its TE properties even at the room temperature.
In this study, the N-impurity bands are treated with both conduction and valence bands taken into account self-consistently.
We found the optimal N concentration per unit cell, $c_{\rm opt}$, which gives the maximum power factor
($PF$) for various temperatures, e.g., $PF=$0.30~$\rm{W/K^2m}$ with $c_{\rm opt}=3.1\times 10^{-5}$ at 300K.
In addition, the electronic thermal conductivity has been estimated, which turn out to be much smaller than the phonon thermal conductivity, leading to the figure of merit as $ZT\sim 0.1$ for N-substituted (20,0) SWCNTs with $c_{\rm opt}=3.1\times 10^{-5}$ at 300K.}

\begin{document}
\maketitle

\section{Introduction}

In 1993, Hicks and Dresselhaus proposed that significant enhancement of the thermoelectric (TE) performance of materials could be realized by employing one dimensional (1D) semiconductors~\cite{rf:hicks}. Single-walled carbon nanotubes (SWCNTs) are of particular interest as high-performance, flexible and lightweight TE 1D materials~\cite{rf:small,rf:nakai,rf:hayashi1,rf:hayashi2,rf:Avery,rf:yanagi,rf:yanagi2,rf:Shimizu,rf:nonoguchi1,rf:nonoguchi2,rf:nonoguchi3,rf:fujigaya1,rf:fujigaya2,rf:fujigaya3,rf:fujigaya4,rf:Jiang,rf:TY-HF01,rf:TY-HF02,rf:blotztrap}.

Both n- and p-type semiconducting SWCNTs are required to develop SWCNT-based TE devices. A great deal of effort has been put into the carrier doping of SWCNTs using various chemical~\cite{rf:nakai,rf:Avery,rf:nonoguchi1,rf:nonoguchi2,rf:nonoguchi3,rf:fujigaya1,rf:fujigaya2,rf:fujigaya3} and field-effect doping methods~\cite{rf:yanagi,rf:yanagi2,rf:Shimizu}.
In the case of field-effect doping, the present authors (T.Y and H.F) have theoretically clarified that an SWCNT exhibits the bipolar TE effect
({\it i.e.,} the sign inversion of Seebeck coefficient from positive (p-type) to negative (n-type) by changing the gate voltage) within the constant-$\tau$ approximation and the self-consistent Born approximation~\cite{rf:TY-HF02}.
On the other hand, in the case of chemical doping, such as with nitrogen (N) and boron (B) doping, the impurity-doped SWCNTs are regarded as strongly disordered systems of which the TE properties cannot, in principle, be theoretically described by the conventional Boltzmann transport theory (BTT). The present authors (T.Y. and H.F.) have recently succeeded in describing the TE properties of N-substituted SWCNTs using the linear response theory (Kubo-L{\" u}ttinger formula~\cite{rf:kubo, rf:Luttinger}) combined with the thermal Green's function technique~\cite{rf:TY-HF01}. In Ref.\citen{rf:TY-HF01}, the authors reported that a decrease in the N concentration of a (10,0) SWCNT increases both the electrical conductivity and the Seebeck coefficient at room temperature ($T=300$~K), and eventually the room-$T$ thermoelectric power factor of the SWCNTs increases {\it monotonically} as the N concentration decreases down to an extremely low concentration of $10^{-5}$ atoms per unit cell. 

In the case of a (10,0) SWCNT with small diameter of $d_{\rm t}=0.78$~nm, the influence of thermal excitation from the valence band to the conduction band on the room-$T$ TE effects is negligible because of the large band gap $E_{\rm g}=0.948$~eV.
On the other hand, when the diameter is larger, the electron-hole excitation probability $(\sim {\rm e}^{-E_{\rm g}/k_{\rm B}T})$ becomes much larger than that for a (10,0) SWCNT. For example, the electron-hole excitation probability at $T=300$~K for a (20,0) SWCNT with a diameter of $d_{\rm t}=1.57$~nm and a band gap of $E_{\rm g}=0.497$~eV, which are a typical diameter and band gap in experiments~\cite{rf:edips}, is much larger ($3.80\times 10^{7}$ times larger) than that for a (10,0) SWCNT. The influence of electron-hole excitation on the TE properties of SWCNTs determines the performance of SWCNT-based TE devices.

To clarify the objective of the present study, we here briefly summarize the above-mentioned two our previous studies~\cite{rf:TY-HF01,rf:TY-HF02}.
In Ref.\citen{rf:TY-HF02}, overall trends of bipolar TE effects have been studied by incorporating the both conduction and valence bands, but the impurity band was not incorporated.
In Ref.\citen{rf:TY-HF01}, we focus on the impurity-band effects on TE properties of (10,0) SWCNTs
with N concentration from $c=10^{-2}$ to $c=10^{-5}$.
Here, we neglect the presence of valence band because the electron-hole excitation probability is negligible even at a high temperature of 400~K (see Appendix~\ref{sec:previous}).
In this situation, the thermoelectric power factor increases
with decreasing the N impurity concentration (see Fig.8 in Ref.\citen{rf:TY-HF01}). On the other hand, for the (20,0) SWCNT, 
the contribution of valence band to the TE effects cannot be neglected even at $\sim$300~K because of the small band gap of $E_{\rm g}=0.497$~eV and it is not clarified yet.
Thus, in this study, we incorporate both the valence and the conduction bands of N-substituted (20,0) SWCNTs
and treat precisely the N-induced impurity band in the band gap using the self-consistent $t$-matrix approximation.
As a result, we found the power factor exhibits the maximum value at a certain concentration of N atoms for a fixed temperature.
In addition, we also estimate the temperature dependence of electronic thermal conductivity $\lambda_{\rm e}$ of N-substituted SWCNTs to be compared to that of phonons 
$\lambda_{\rm ph}$ and then estimate the figure of merit, $ZT$.

\section{Theoretical Modeling and Formulation~\label{sec:2}}
\subsection{Linear Response Theory for Thermoelectric Effects~\label{sec:2.1}}
In the presence of both an electric field $\mathcal{E}$ and a temperature gradient of $dT/dz$ along the $z$-direction in a material 
({\it e.g.}, the tube axis of an SWCNT), the electrical current density $J$ is generally given by
\begin{eqnarray}
J=L_{11}\mathcal{E}-\frac{L_{12}}{T}\frac{dT}{dz}
\label{eq:J}
\end{eqnarray}
within the linear response with respect to $\mathcal{E}$ and $dT/dz$~\cite{rf:notation}. Here, $L_{11}$ and $L_{12}$ are the electrical conductivity and the thermoelectrical conductivity, respectively.
Using $L_{11}$ and $L_{12}$, the Seebeck coefficient $S$ is expressed as
\begin{eqnarray}
S=\frac{1}{T}\frac{L_{12}}{L_{11}}
\label{eq:S}
\end{eqnarray}
and the power factor $PF$, which is one of the figures of merit for TE materials, is described by
\begin{eqnarray}
PF\equiv L_{11} S^2=\frac{1}{T^2}\frac{L_{12}^2}{L_{11}}.
\label{eq:PF}
\end{eqnarray}
The expression of $L_{11}$ and $L_{12}$ is given by
\begin{eqnarray}
L_{11}&=&\int_{-\infty}^\infty \!\!\!dE\left(-\frac{\partial f(E-\mu)}{\partial E}\right)\alpha(E),
\label{eq:L11}\\
L_{12}&=&-\frac{1}{e}\int_{-\infty}^\infty \!\!\!dE\left(-\frac{\partial f(E-\mu)}{\partial E}\right)(E-\mu)\alpha(E)
\label{eq:L12}
\end{eqnarray}
in terms of the spectral conductivity $\alpha(E)$.
Here, $e$ is the elementary charge,
$\mu$ is the chemical potential and $f(E-\mu)=1/(\exp((E-\mu)/k_{\rm B}T)+1)$ is the Fermi-Dirac distribution function. 
$S$ in Eq.~(\ref{eq:S}) and $PF$ in Eq.~(\ref{eq:PF}) can thus be determined from 
Eqs.~(\ref{eq:L11}) and (\ref{eq:L12}) once $\alpha(E)$ is known.
To the best of our knowledge, the expression of $L_{11}$ and $L_{12}$ in Eqs.~(\ref{eq:L11}) and (\ref{eq:L12}) was first 
proposed by Sommerfeld and Bethe in 1933~\cite{rf:sommerfeld}, subsequently by Mott and Jones~\cite{rf:mott1936}, and then by Wilson~\cite{rf:wilson}.
Recently, the authors (T.Y. and H.F.) applied the Sommerfeld-Bethe (SB) relation expressed as Eqs.~(\ref{eq:L11}) and (\ref{eq:L12}) to disordered N-substituted SWCNTs using a simple tight-binding model combined with a self-consistent $t$-matrix approximation~\cite{rf:TY-HF01}.
Also, Akai and co-workers adopted the SB relation to treat the disordered metal alloys using the density functional theory combined with a coherent potential approximation (CPA) ~\cite{rf:akai-kkr1,rf:akai-kkr2}.
More recently, Ogata and Fukuyama clarified the range of validity of the SB relation, even for correlated systems including electron-phonon coupling and electron correlations~\cite{rf:ogata-fukuyama2019}.

\subsection{Effective-Mass Hamiltonian of SWCNTs~\label{sec:2.2}}
In this subsection, we briefly review the electronic structure of semiconducting SWCNTs with zigzag-type edges (z-SWCNTs).
In our previous paper in Ref.\citen{rf:TY-HF02},  we gave a one-dimensional Dirac Hamiltonian with an energy dispersion
\begin{eqnarray}
\epsilon_{k,q}^{(\pm)}=\pm\sqrt{(\hbar v_qk)^2+\Delta_q^2}
\label{eq:Dirac_disp}
\end{eqnarray}
for the effective Hamiltonian of semiconducting $(n,0)$ SWCNTs near the conduction ($+$) and valence ($-$) band edges,
where $k$ is the wavenumber along the tube-axial direction and
$q$ specifies the two pairs of lowest-conduction and highest-valence bands:
\begin{eqnarray}
q=\left\{
\begin{array}{l}
q_1\equiv(2n+1)/3\\
q_2\equiv(4n-1)/3,
\end{array}
\quad {\rm for}\; n\; {\rm mod}\; 3=1
\right.
\end{eqnarray}
and
\begin{eqnarray}
q=\left\{
\begin{array}{l}
q_1\equiv(2n-1)/3\\
q_2\equiv(4n+1)/3,
\end{array}
\quad {\rm for}\; n\; {\rm mod}\; 3=2.
\right.
\end{eqnarray}
$\Delta_q$ is a half of the band gap ({\it i.e.}, $E_{\rm g}\equiv 2\Delta_q$) and $v_q$ is a velocity, expressed as
\begin{eqnarray}
\Delta_q=\gamma_0\left|1+2{\rm cos}\left(\frac{\pi q}{n}\right)\right|
\label{eq:}
\end{eqnarray}
and
\begin{eqnarray}
v_q=-\frac{a_z\gamma_0}{\hbar}{\rm cos}\frac{\pi q}{n}
\label{eq:}
\end{eqnarray}
where $\gamma_0=2.7$~eV is the hopping integral between nearest-neighbor carbon atoms and
$a_z=0.426$~nm is the unit-cell length for an $(n,0)$ SWCNT~\cite{rf:hamada,rf:saito}.
The energy origin ($E=0$~eV) in Eq.~(\ref{eq:Dirac_disp}) is set at the middle of the band gap, $E_g$.
In the small-$k$ region that obeys $k^2\ll(\Delta_q/\hbar v_q)^2$, the energy dispersion in Eq.~(\ref{eq:Dirac_disp}) is reduced to
\begin{eqnarray}
\epsilon_{k,q}^{(\pm)}=\pm\left(\frac{\hbar^2k^2}{2m_q^*}+\Delta_q\right)
\label{eq:k2-dispersion}
\end{eqnarray}
with the effective mass $m_q^*=\Delta_q/v_q^2$ for both conduction and valence bands.
Thus, the effective Hamiltonian is also given by
\begin{eqnarray}
\mathcal{H}_0=\sum_{k}\left(\epsilon_{k}^{(+)}c_{k}^{\dagger} c_{k}+\epsilon_{k}^{(-)}d_{k}^{\dagger} d_{k}\right),
\label{eq:H_eff}
\end{eqnarray}
with Eq.~(\ref{eq:k2-dispersion}), where $c_{k}^{\dagger}$ and $d_{k}^{\dagger}$ ($c_{k}$ and $d_{k}$) are the creation (annihilation) operators for the conduction and valence band electrons, respectively. The spin and orbital degrees of freedom $q$ are omitted from Eq.~(\ref{eq:H_eff}).

At this point, we take account of the random potential term in $\mathcal{H}_0$ in Eq.~(\ref{eq:H_eff}) such that
\begin{eqnarray}
\mathcal{H}=\mathcal{H}_0+V_0\sum_{\langle j\rangle} c_j^\dagger c_j
\label{eq:H_eff+V0}
\end{eqnarray}
to examine the effects of N-doping on SWCNTs. Here, $V_0$ is the attractive potential ($V_0<0$) for an N atom in an SWCNT.
For example, $V_0=-0.91$~eV for (20,0) SWCNTs (see Sec.~\ref{sec:2.3} for details).
In Eq.~(\ref{eq:H_eff+V0}), $c_{j}^\dagger$ ($c_{j}$) is 
the creation (annihilation) operator of an electron at the $j$th impurity position, and ${\langle j\rangle}$ represents the sum with respect to randomly distributed impurity positions for a fixed average concentration of $c=N_{\rm imp}/N_{\rm unit}$, where $N_{\rm imp}$ is the total number of 
impurity positions and $N_{\rm unit}$ is the number of unit cells in a pristine SWCNT with the length $L$. 

We also confirm that the small-$k$ condition of $|\hbar vk|\ll \Delta$ is satisfied within the temperature region of 0~$<T<$~ 500K discussed in this paper.

\subsection{Self Energy due to Impurity Potential~\label{sec:2.3}}
\begin{figure}[t]
  \begin{center}
  \includegraphics[keepaspectratio=true,width=80mm]{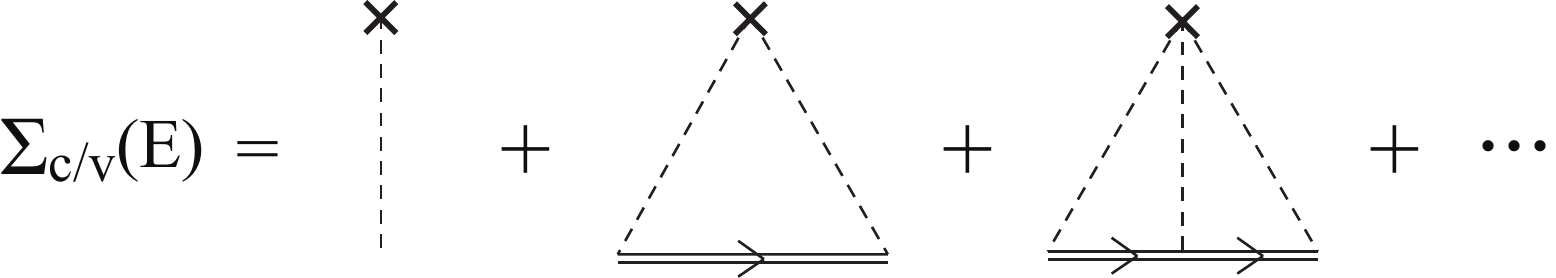}
  \end{center}
\caption{Diagram of a self-consistent $t$-matrix approximation for the retarded self-energy.
The crosses, dashed lines and solid double lines with arrows respectively denote the impurity sites, the impurity potential and 
the one-particle retarded Green's function to be determined self-consistently.}
\label{fig:01}
\end{figure}

The modification of thermoelectric effects by randomly distributed impurities will be studied based on the thermal Green's function 
formalism through self-energy corrections of the Green's functions. In this study, the influence of random N potential on the conduction- and
valence-band electrons is incorporated into the retarded self energy using the self-consistent $t$-matrix approximation as shown in 
Fig.~\ref{fig:01}~\cite{rf:TY-HF01,rf:saitoh_HF-YU-HS,rf:ogata-fukuyama,rf:matsuura},
which corresponds to the dilute limit of CPA for binary alloys~\cite{rf:cpa1,rf:cpa2,rf:akai-kkr1,rf:akai-kkr2}.

Within the self-consistent $t$-matrix approximation,
both the self energies $\Sigma_{\rm c/v}(E)$ for conduction/valence-band electrons in an N-substituted SWCNT are independent of $k$ because of 
the short-range of the impurity potential in Eq.~(\ref{eq:H_eff+V0}) and are determined by the requirement of self-consistency, as
\begin{eqnarray}
\Sigma_{\rm c/v}(E)=\frac{cV_0}{1-X_{\rm c/v}(E)}, \quad {\rm Im}~\Sigma_{\rm c/v}(E)<0
\label{eq:retarded_self-energy}
\end{eqnarray}
with 
\begin{eqnarray}
X_{\rm c/v}(E)=\frac{V_0}{N_{\rm unit}}\sum_{k}\frac{1}{E-\epsilon_{k}^{(\pm)}-\Sigma_{\rm c/v}(E)},
\label{eq:XE0}
\end{eqnarray}
where $\pm$ corresponds to c/v, respectively.
The $k$-summation in Eq.~(\ref{eq:XE0}) can be analytically performed by substituting Eq.~(\ref{eq:k2-dispersion}) into Eq.~(\ref{eq:XE0}), and we obtain
\begin{eqnarray}
X_{\rm c/v}(x)=\mp\frac{i}{2}\frac{v_0}{\sqrt{\pm(x-\sigma_{\rm c/v}(x))-\delta}},
\label{eq:XE}
\end{eqnarray}
where ${\rm Im}\sqrt{\pm(x-\sigma_{\rm c/v}(x))-\delta}>0$, and $x\equiv E/t$, $v_0\equiv V_0/t$, $\delta\equiv \Delta/t$ and 
$\sigma_{\rm c/v}\equiv\Sigma_{\rm c/v}/t$ with
\begin{eqnarray}
t\equiv \frac{\hbar^2}{2m^*a_z^2}
\label{eq:t_ch}
\end{eqnarray}
equal to the characteristic energy of an SWCNT. 
From Eqs.~(\ref{eq:retarded_self-energy}) and (\ref{eq:XE}), the self-consistent equation for $\sigma_{\rm c/v}(x)$ is given by
\begin{eqnarray}
\sigma_{\rm c/v}(x)=\frac{cv_0}{1\pm\frac{i}{2}\frac{v_0}{\sqrt{\pm(x-\sigma_{\rm c/v}(x))-\delta}}}.
\label{eq:self-consistent_Sigma}
\end{eqnarray}
Equation~(\ref{eq:self-consistent_Sigma}) can also be rewritten as 
\begin{eqnarray}
x=\sigma_{\rm c/v}\pm\delta\mp\frac{v_0^2\sigma_{\rm c/v}^2}{4(\sigma_{\rm c/v}-cv_0)^2}
\label{eq:x}
\end{eqnarray}
or as the cubic equation for $\sigma_{\rm c/v}$,
\begin{eqnarray}
a_3\sigma_{\rm c/v}^3(x)+a_2\sigma_{\rm c/v}^2(x)+a_1\sigma_{\rm c/v}(x)+a_0=0,
\label{eq:third_Sigma}
\end{eqnarray}
with $a_3=1$, $a_2=-(x\mp \delta+2cv_0\pm v_0^2/4)$, $a_1=cv_0\{2(x\mp \delta)+cv_0\}$, and $a_0=-(x\mp \delta)(cv_0)^2$, where
the upper/lower sign is for the conduction/valence-band electrons.
Equation~(\ref{eq:third_Sigma}) indicates that for each energy, $x=E/t$, there are three solutions of $\sigma_{\rm c/v}(x)$:
three real solutions or one real solution and two complex solutions. 
\begin{figure}[t]
  \begin{center}
  \includegraphics[keepaspectratio=true,width=80mm]{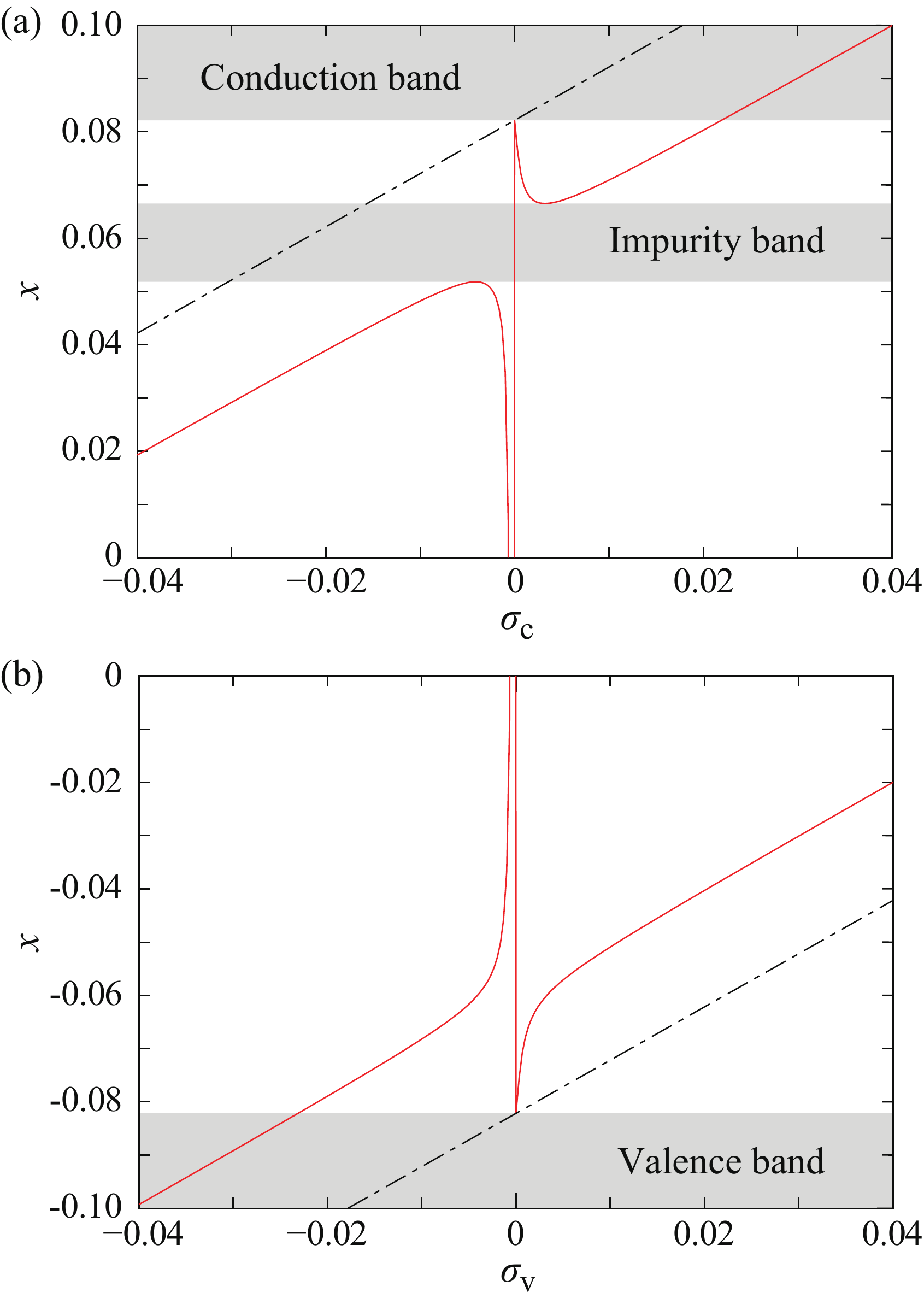}
  \end{center}
\caption{(Color online) The $x$-$\sigma_{\rm c/v}$ relations for (a) conduction-band electrons and (b) valence-band electrons of (20,0) SWCNT with $c=10^{-3}$. 
The shaded regions indicate the energy region where the cubic equation in Eq.~(\ref{eq:third_Sigma}) has one real solution and two complex solutions. 
The broken lines denotes $x=\pm\delta+\sigma_{\rm c/v}$.}
\label{fig:0X}
\end{figure}

Figures~2(a) and 2(b) show a real $x$ as a function of real $\sigma_{\rm c/v}$ for (20,0) SWCNT with $c=10^{-3}$. 
In the shaded regions of $x$, there are complex solutions of $\sigma_{\rm c/v}(x)$.
The boundaries between the finite- and zero-DOS regions [{\rm e.g.}, the conduction-band bottom $E_{\rm c}$ and 
the valence-band top $E_{\rm v}$] can be determined using the condition $dx/d\sigma_{\rm c/v}=0$. 
In addition, we can see in Fig.~2(a) that the impurity band appears just below the conduction band.
In the limit of $c\to 0$, the binding energy of bound state can be calculated from a pole of $t$-matrix ${\mathcal T}_{\rm c}(E)=V_0/(1-X_{\rm c}(E))$ for
the conduction-band electron as
\begin{eqnarray}
E_{\rm b}=t\left(\frac{v_0}{2}\right)^2.
\label{eq:Eb-V0}
\end{eqnarray}
Thus, once $E_{\rm b}$ is given, the attractive potential $V_0$ can be determined by Eq.~(\ref{eq:Eb-V0}).
For the (20,0) SWCNT, $E_{\rm b}$ is known to be $E_{\rm b}=0.068$~eV~\cite{rf:koretsune},
and eventually the attractive potential is $V_0=-0.91$~eV. In other words, the single impurity level is located at $E=0.18$~eV ($x=0.060$).

Note that, in CPA methods, including the present self-consistent $t$-matrix approximation, the spectral conductivity
$\alpha(E)$ becomes finite once DOS becomes finite (see Sec.~\ref{sec:alpha}), since CPA ignores the effects of Anderson localization due to 
the interference effects of scattered waves, which can lead to finite DOS even in the energy region where the conductivity is zero. 
It is known that every state is localized in one and two dimensions in the presence of finite scattering~\cite{rf:nagaoka}. However, once the system size or 
temperature becomes finite, the effects of Anderson localization are greatly reduced. This situation is assumed in the present study 
and hence the band edges in the CPA are used to represent the effective mobility edges.

\subsection{Density of States $\rho$}
\begin{figure}[t]
  \begin{center}
  \includegraphics[keepaspectratio=true,width=80mm]{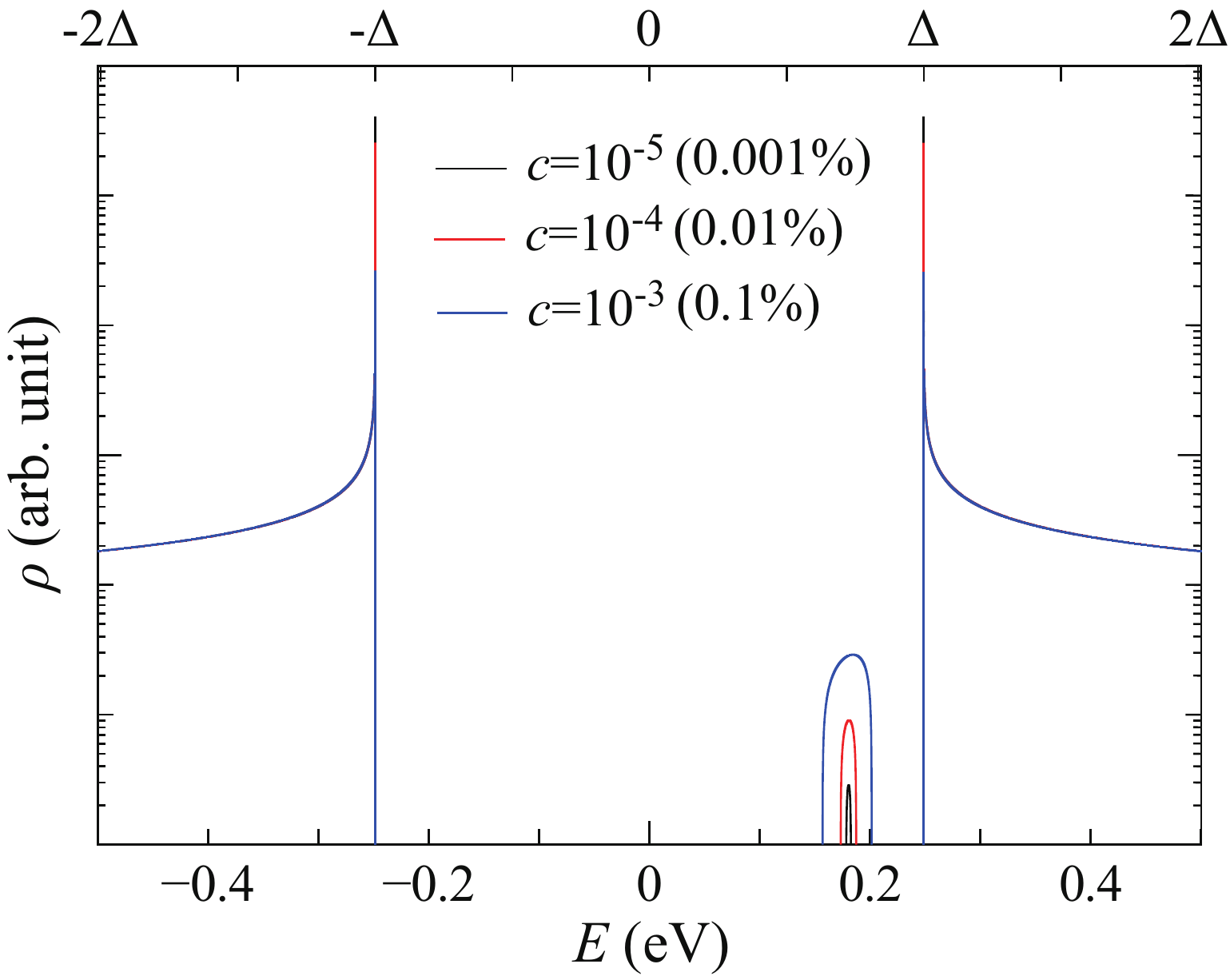}
  \end{center}
\caption{(Color online) Density of states (DOS) results for an N-substituted (20,0) SWCNT for which $c=10^{-3}$ (blue curve), 
$10^{-4}$ (red curve) and $10^{-5}$ (black curve) in a logarithmic scale.  
The energy origin ($E=0$~eV) was set equal to the middle of the band gap of the pristine (20,0) SWCNT.}
\label{fig:03}
\end{figure}

Once the self energies $\Sigma_{\rm c}(E)$ and $\Sigma_{\rm v}(E)$ are obtained via the above procedure, the density of states (DOS) 
can be determined as follows. Within the present approximations, the DOS per the unit cell for each spin and each orbital consists of two parts, as
\begin{eqnarray}
\rho(E)=\rho_{\rm c}(E)+\rho_{\rm v}(E),
\label{rf:dos_cv}
\end{eqnarray}
where $\rho_{\rm c}(E)$ is the DOS including the contribution from the conduction and impurity bands, 
and $\rho_{\rm v}(E)$ is the DOS of the valence band, which are given by
\begin{eqnarray}
\rho_{\rm c/v}(E)&=&-\frac{1}{\pi V_0}{\rm Im}~X_{\rm c/v}(E)\label{eq:dos_dif}
\label{eq:DOS0}\\
&=&\pm\frac{1}{2\pi \sqrt{t}}{\rm Re}\frac{1}{\sqrt{\pm\left(E-\Sigma_{\rm c/v}(E)\right)-\Delta}},
\label{eq:DOS}
\end{eqnarray}
where ${\rm Im}\sqrt{\pm\left(E-\Sigma_{\rm c/v}(E)\right)-\Delta}>0$.
The signs + and - correspond to $\rho_{\rm c}$ and $\rho_{\rm v}$, respectively.

Equation~(\ref{eq:DOS}) indicates that for the region of $x$ with complex solutions of $\sigma_{c/v}(x)$, the DOS will be finite, i.e., 
in the shaded region in Figs.~2(a) and 2(b). One caution is that the DOS can be finite even for real  $\sigma_{c/v}(E)$ if 
$\pm\left(E-\Sigma_{\rm c/v}(E)\right)-\Delta>0$ ({\rm i.e.}, $\pm(x-\sigma_{\rm c/v})-\delta>0$); however all the solutions of $\sigma_{\rm c/v}(x)$ 
satisfying Eq.~(\ref{eq:x}) are in the region of $\pm(x-\sigma_{\rm c/v})-\delta<0$ resulting in the absence of DOS [see also Figs.~2(a) and 2(b)].

Figure~\ref{fig:03} presents the DOS for (20,0) SWCNTs with various concentrations of N atoms ($c=10^{-3}$ (0.1\%), 
$10^{-4}$ (0.01\%) and $10^{-5}$ (0.001\%)). For a (20,0) SWCNT, $E_{\rm g}=$0.497~eV ($\Delta=$0.248~eV) and $m^*$=0.069$m_0$ are used, 
where $m_0=9.11\times10^{-31}$kg is the electron mass in a vacuum. Figure~\ref{fig:03} shows that the N impurity band appears around $E=\Delta-E_{\rm b}$($=0.18$~eV),
and both the impurity band width and height increase with $c$. The DOS for the pristine (20,0) SWCNT without any defects or impurities exhibits
a van Hove singularity at the conduction band bottom of $E=\Delta$($=0.248$~eV) and the valence band top of $E=-\Delta$.
Although the van Hove singularity disappears in the presence of N impurities, the DOS results in Fig.~\ref{fig:03} have a sharp peak near $E=\pm\Delta$, which implies that the electrons in the conduction and valence electronic states are not significantly disordered by N impurities.
We can see that as $c$ increases, the DOS near $E=\pm\Delta$ decreases strongly in comparison with that in the conduction and the valence bands.

\subsection{Chemical Potential $\mu$~\label{sec:2.5}}
\begin{figure}[t]
  \begin{center}
\includegraphics[keepaspectratio=true,width=80mm]{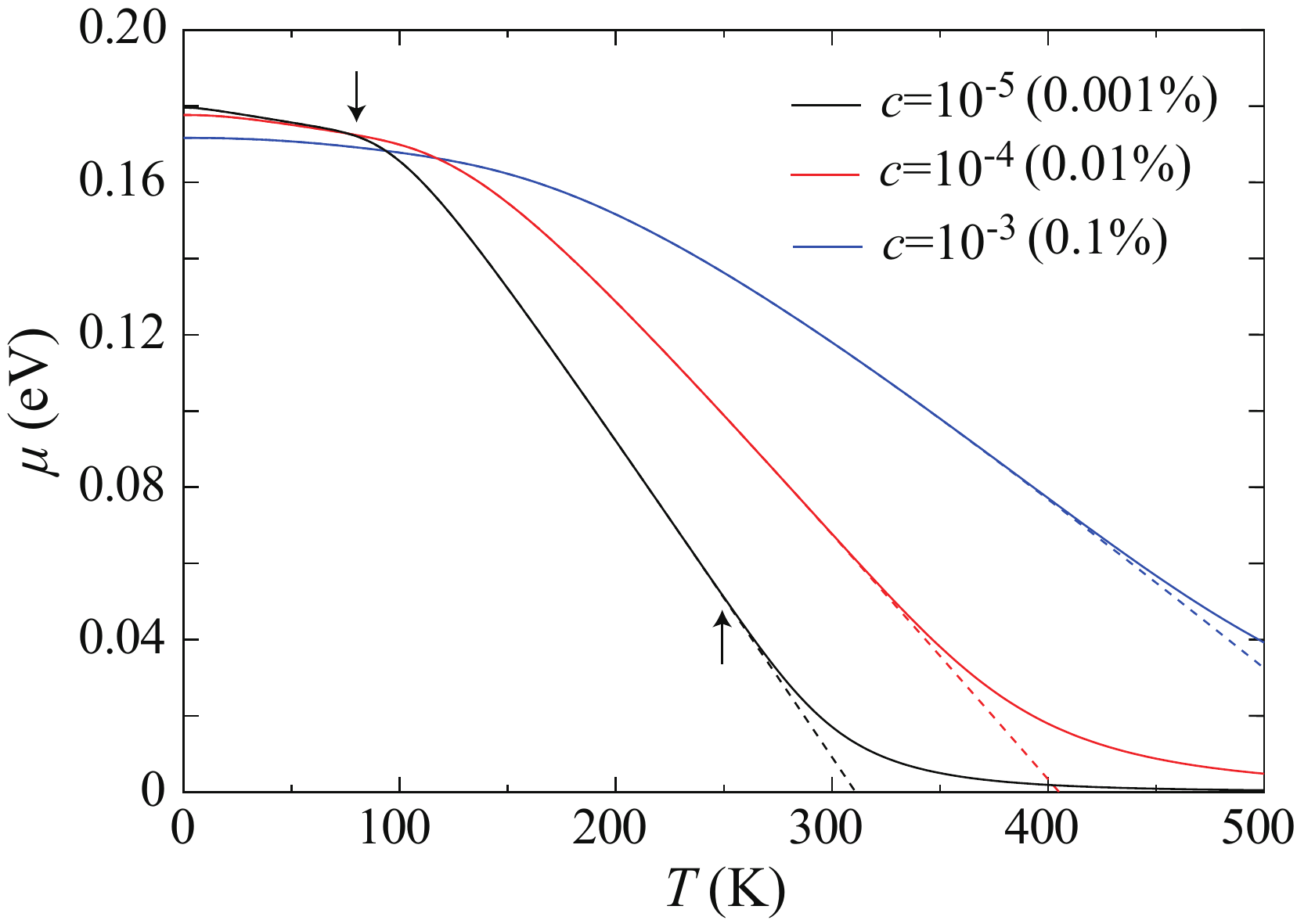}
  \end{center}
\caption{(Color online) Temperature dependence of the chemical potentials $\mu(T)$ of N-substituted (20,0) SWCNTs for which $c=10^{-3}$ 
(blue solid curve), $10^{-4}$ (red solid curve) and $10^{-5}$ (black solid curve). The dashed curves are chemical potential curves where 
the valence band is not taken into account in the calculation. The arrows indicate $T$=80 K and 250 K.}
\label{fig:mu}
\end{figure}

At $T=0$, the chemical potential (Fermi energy) lies in the impurity band.
We now explain how to determine the $T$-dependence of the chemical potential $\mu(T)$.
Once the DOS in Eq.~(\ref{eq:DOS}) is obtained,
the $T$-dependence of the chemical potential $\mu(T)$ can be determined 
with respect to the total electron density:
\begin{eqnarray}
4\int_{-\infty}^{\infty}\rho(E)f(E-\mu(T))dE=4\int_{-\infty}^{-|E_{\rm v}|}\rho_{\rm v}(E)dE+c,
\label{eq:DOS2}
\end{eqnarray}
where the left and right hand sides indicate the total amount of carriers per unit cell of the system at finite $T$ and zero $T$, respectively.
The factor $4$ originates from the spin and orbital degeneracy of z-SWCNTs and $E_{\rm v}=-|E_{\rm v}|$ is the valence-band top, which
can be determined by the condition $dx/d\sigma_{\rm v}=0$.

Figure~\ref{fig:mu} presents the $T$-dependence of $\mu(T)$ for N-substituted (20,0) SWCNTs with $c=10^{-3}$ (blue solid curve), $10^{-4}$ (red solid curve) 
and $10^{-5}$ (black solid curve). The dashed curves are $\mu(T)$ where the valence band is not taken into account in the calculation,
which was previously discussed in Ref.\citen{rf:TY-HF01}.   We now focus on the case of $c=10^{-5}$ (black solid curve) as an example.
The black solid curve shows characteristic changes around $T\sim80$~K and $T\sim250$~K, as indicated by the arrows.
In the ionization region of $T\lesssim80$~K (see Appendix~\ref{sec:n_c}), $\mu(T)$ lies in the impurity band and decreases slowly 
with an increase in $T$. As $T$ increases further, the system shows a crossover from the ionization region to the exhaustion region (see Appendix~\ref{sec:n_c}).
In this crossover region of 80 K$\lesssim T\lesssim$250 K, $\mu(T)$ decreases rapidly. Over $T\sim250$K, the black solid curve deviates upward from 
the black dashed curve and approaches the center of the band gap ($E=0$) in the high-$T$ limit because the valence band electrons begin to be thermally excited
from the valence band to the conduction band. This temperature region, $T\gtrsim 250$K, is the so-called intrinsic region (see Appendix~\ref{sec:n_c}).
Similar features are evident in the red and blue solid curves. It should be noted that the two characteristic temperatures indicated by the arrows in Fig.~\ref{fig:mu} 
shift toward higher $T$ as $c$ increases.

\subsection{Spectral Conductivity $\alpha(E)$~\label{sec:alpha}}
\begin{figure}[t]
  \begin{center}
\includegraphics[keepaspectratio=true,width=80mm]{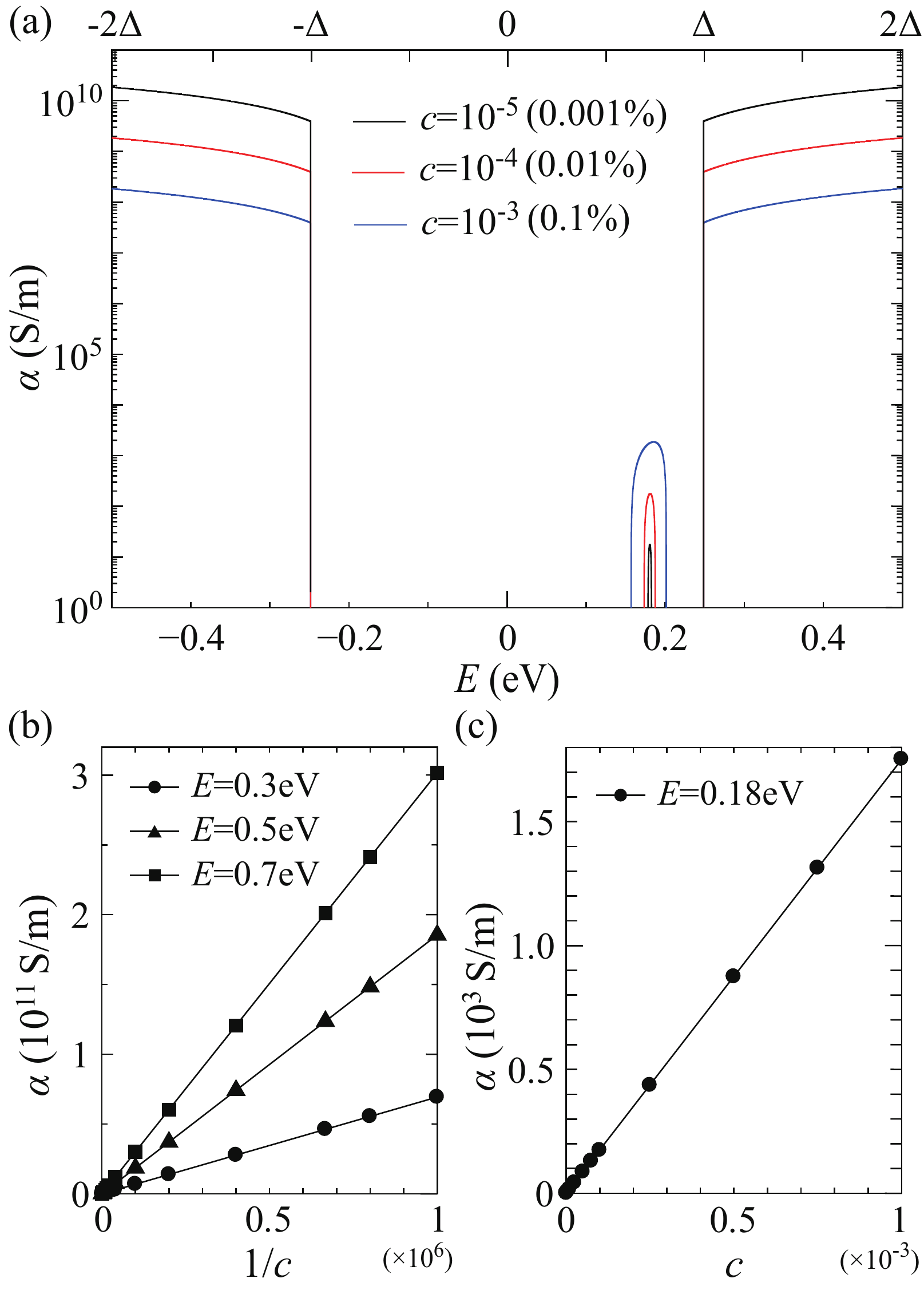}
  \end{center}
\caption{(Color online) (a) Spectral conductivity $\alpha$ for N-substituted (20,0) SWCNTs with $c=10^{-3}$ (blue curve), $10^{-4}$ (red curve) 
and $10^{-5}$ (black curve). The energy origin ($E=0$~eV) was set equal to the middle of the band gap of the pristine (20,0) SWCNTs.
(b) $c$-dependence of $\alpha$ at $E=0.3$~eV (circles), $E=0.5$~eV (triangles) and $E=0.7$~eV (squares).
(c) $c$-dependence of $\alpha$ at $E=\Delta-E_{\rm b}(=0.18$~eV).}
\label{fig:alpha}
\end{figure}

Similar to the expression of the DOS in Eq.~(\ref{rf:dos_cv}), the spectral conductivity $\alpha(E)$ can also be divided into two parts, as
\begin{eqnarray}
\alpha(E)=\alpha_{\rm c}(E)+\alpha_{\rm v}(E)
\label{eq:alpha_E}
\end{eqnarray}
within the present approximation. Here, $\alpha_{\rm c}(E)$ is the spectral conductivity of conduction and impurity band electrons, and 
$\alpha_{\rm v}(E)$ is the spectral conductivity of valence band electrons. $\alpha_{\rm c}(E)$ and $\alpha_{\rm v}(E)$ are 
given by Refs.~\citen{rf:Jonson_1980,rf:TY-HF01, rf:TY-HF02} \begin{eqnarray}
\alpha_{\rm c/v}(E)=4\frac{e^2\hbar}{\pi V}\sum_{k} v_{k}^2\left[{\rm Im~} G_{\rm c/v}(k, E)\right]^2,
\label{eq:sc_cv}
\end{eqnarray}
where the factor 4 comes from the spin degeneracy and the orbital degeneracy of z-SWCNTs~\cite{rf:hamada,rf:saito},
$v_{k}$ is the group velocity of an electron with wavenumber $k$,
and $V$ is the volume of the system. Here, $G_{\rm c/v}(k, E)$ is the retarded Green's function
\begin{eqnarray}
G_{\rm c/v}(k, E)=\frac{1}{E-\epsilon^{(\pm)}_{k}-\Sigma_{\rm c/v}(E)}.
\label{eq:retarded_G}
\end{eqnarray}
Furthermore, within the effective-mass approximation for z-SWCNTs in Eq.~(\ref{eq:k2-dispersion}), 
the $k$-summation in Eq.~(\ref{eq:sc_cv}) can be performed analytically and $\alpha_{\rm c/v}(x)$ are given by
\begin{eqnarray}
\alpha_{\rm c/v}(x)=\frac{2e^2}{\pi\hbar}\frac{a_z}{A}\frac{\left({\rm Re}\sqrt{\pm\left(x-\sigma_{\rm c/v}\right)-\delta}\right)^2}{\left|\pm\left(x-\sigma_{\rm c/v}\right)-\delta\right|{\rm Im}\sqrt{\pm\left(x-\sigma_{\rm c/v}\right)-\delta}},
\label{eq:alpha_eff-m}
\end{eqnarray}
where the signs $+$ and $-$ correspond to $\alpha_{\rm c}$ and $\alpha_{\rm v}$, respectively,
and $A$ is the cross-sectional area of an SWCNT ($A\equiv\pi d_{\rm t}\delta_w$ is conventionally used as the effective cross-sectional area of an SWCNT, 
where $d_{\rm t}=1.57$ nm is the diameter of a (20,0) SWCNT and $\delta_w=0.34$ nm is the van der Waals diameter of carbon).

Figure~\ref{fig:alpha}(a) shows $\alpha(E)$ for N-substituted (20,0) SWCNTs with various concentrations of N impurities ($c=10^{-3}$ (blue solid curve), 
$10^{-4}$ (red solid curve) and $10^{-5}$ (black solid curve)). $\alpha(E)$ has finite value once DOS is finite.
With a decrease in $c$, $\alpha(E)$ in the energy regions of $E\ge E_{\rm c}$ and $E\le E_{\rm v}$ is proportional to $1/c$, as shown in Fig.~\ref{fig:alpha}(b).
This can be understood by the BTT expression $\alpha_{\rm c/v}(E)\propto \tau_{\rm c/v}$ with
the relaxation time $\tau_{\rm c/v}$. Since $\tau_{\rm c/v}$ is proportional to $1/c$ within the $t$-matrix approximation,
we obtain $\alpha_{\rm c/v}(E)\propto 1/c$ (see Appendix~\ref{sec:t_mat}).
In contrast to the conduction/valence-band energy region, $\alpha_{\rm c}(E)$ for the impurity-band energy region, which cannot be described by the BTT, is proportional to $c$, 
as shown in Fig.~\ref{fig:alpha}(c). This is because that the averaged distance of N impurities becomes short in proportion to $c$.

\section{Numerical Results and Discussion~\label{sec:3}}
\subsection{Electrical Conductivity $L_{11}$~\label{sec:3.2.1}}
\begin{figure}[t]
 \begin{center}
\includegraphics[keepaspectratio=true,width=80mm]{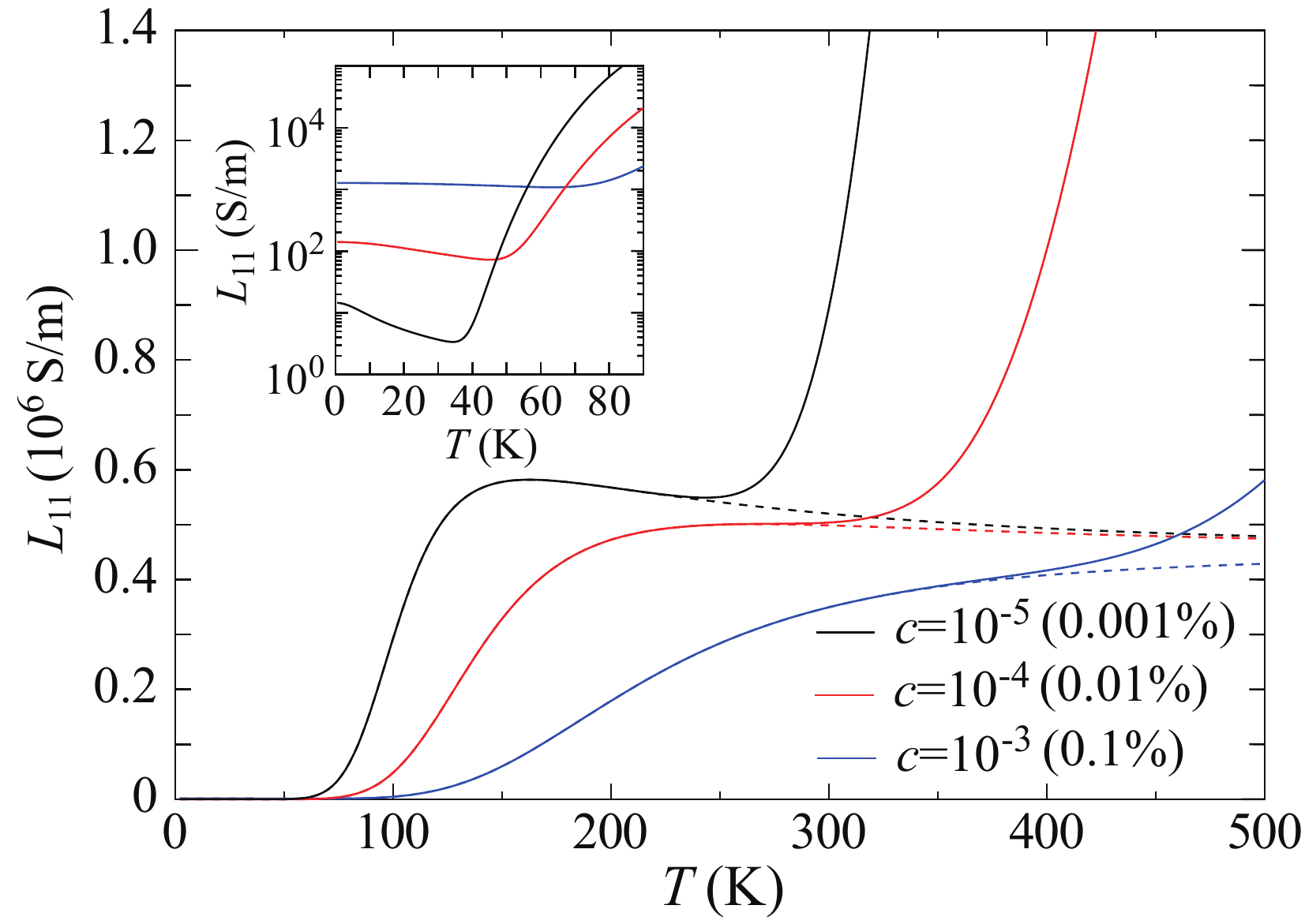}
  \end{center}
\caption{(Color online)
Temperature dependence of the electrical conductivity $L_{\rm 11}$ of N-substituted (20,0) SWCNTs with $c=10^{-3}$ (blue solid curve), 
$10^{-4}$ (red solid curve) and $10^{-5}$ (black solid curve). The dashed curves are $L_{\rm 11}$ where the valence band is not taken 
into account in the calculation.
}
\label{fig:L11}
\end{figure}

We now discuss the $T$-dependence of $L_{\rm 11}$ for N-substituted (20,0) SWCNTs, which can be calculated by the substitution of Eq.~(\ref{eq:alpha_eff-m}) 
into Eq.~(\ref{eq:L11}). Figure~\ref{fig:L11} shows the $T$-dependence of $L_{\rm 11}$ for $c=10^{-3}$ (blue solid curve), $10^{-4}$ (red solid curve) and 
$10^{-5}$ (black solid curve). The dashed curves are $L_{\rm 11}$ where the valence band is not taken into account in the calculation~\cite{rf:TY-HF01}.
Here, we focus on $L_{\rm 11}$ for $c=10^{-5}$ (black solid curve) as an example. In Fig.~\ref{fig:L11}, the black solid curve exhibits two rapid increases 
at $T\sim 40$ K (see also the inset of Fig.~\ref{fig:L11}) and $T\sim 250$ K. The increase at $T\sim 40$ K originates from the change in the transport regime of 
this system from impurity-band conduction to conduction-band conduction. On the other hand, at $T\sim 250$~K where electrons begin to 
be excited from the valence band to the conduction band, the black solid curve begins to deviate upward from the dashed curve. This is because the valence-band 
holes contribute to $L_{\rm 11}$ in addition to the conduction band electrons at $T\gtrsim250$ K. In the intermediate temperature region of 
170 K$\lesssim T\lesssim$250 K, which corresponds to the exhaustion region, the conduction band electron density is almost constant with $T$, as shown 
in Appendix~\ref{sec:n_c}, and therefore the $T$-dependence of $L_{\rm 11}$ is weak. The $T$-dependence of $L_{\rm 11}$ in the exhaustion region is 
discussed in Appendix~\ref{sec:exhaustion}.

In the last part of this section, we consider the other solid curves (the red and blue solid curves in Fig.~\ref{fig:L11}) to clarify the $c$-dependence of $L_{\rm 11}$.
In the extremely low-$T$ region shown in the insets of Fig.~\ref{fig:L11}, $L_{\rm 11}$ increases with $c$, in contrast to the high-$T$ $L_{\rm 11}$.
This is due to the opposite tendency of the $c$-dependency of $\alpha_{\rm c}(E)$ for the conduction-band and impurity-band energy regions
 [see Figs.~\ref{fig:alpha}(b) and~\ref{fig:alpha}(c)].

\begin{figure}[t]
 \begin{center}
\includegraphics[keepaspectratio=true,width=80mm]{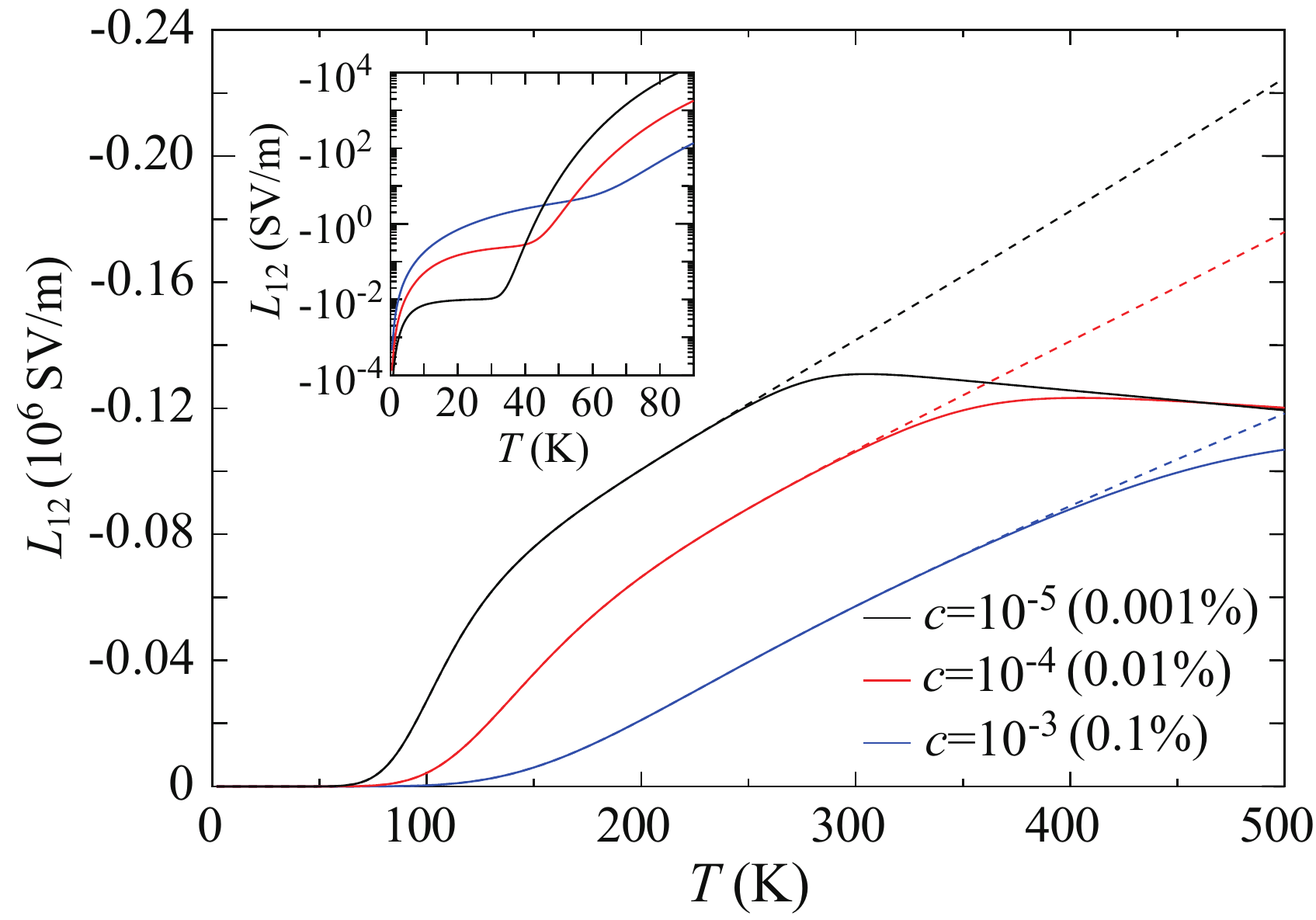}
  \end{center}
\caption{(Color online) Temperature dependence of the electrical conductivity $L_{\rm 12}$ of N-substituted (20,0) SWCNTs with 
$c=10^{-3}$ (blue solid curve), $10^{-4}$ (red solid curve) and $10^{-5}$ (black solid curve). The dashed curves are $L_{\rm 12}$ where 
the valence band is not taken into account in the calculation. Note that the vertical axis is negative.}
\label{fig:L12}
\end{figure}

\subsection{Thermoelectrical Conductivity $L_{\rm 12}$~\label{sec:3.2.1}}
In this section, we discuss the $T$-dependence of $L_{\rm 12}$ for N-substituted (20,0) SWCNTs, which can be calculated by the substitution of 
Eq.~(\ref{eq:alpha_eff-m}) into Eq.~(\ref{eq:L12}). Figure~\ref{fig:L12} shows the $T$-dependence of the $L_{\rm 12}$ value of N-substituted 
(20,0) SWCNTs with $c$=$10^{-3}$ (blue solid curve), $10^{-4}$ (red solid curve) and $10^{-5}$ (black solid curve). The dashed curves are $L_{\rm 12}$ 
where the valence band is not taken into account in the calculation. Here, we focus on the case of $c=10^{-5}$ (black solid curve) as an example.
In Fig.~\ref{fig:L12}, the black solid curve shows a rapid increase at $T\sim 30$ K (see also the inset of Fig.~\ref{fig:L12}) and deviates downward from 
the black dashed curve at $T\sim 250$ K. The rapid increase at $T\sim30$ K is due to the contribution to $L_{\rm 12}$ from 
the conduction band electrons becoming more dominant than that from the impurity band electrons. It should be noted that the crossover temperature 
($T\sim 30$ K) of $L_{\rm 12}$ is lower than that of $L_{\rm 11}$ ($T\sim 40$ K) shown by the black solid curve in the inset of Fig.~\ref{fig:L11}.
This difference implies that $L_{\rm 12}$ is more sensitive to the thermal excitation of carriers than $L_{\rm11}$, and the difference determines 
the low-$T$ behavior of the Seebeck coefficient, as explained in Sec.~\ref{sec:S}. On the other hand, the deviation of the black solid curve from 
the dashed curve at $T\sim250$ K is due to cancellation between the contributions from conduction band electrons and valence band holes to $L_{\rm 12}$.
In addition, we discuss the $T$-dependence of $L_{\rm 12}$ in the intermediate $T$ region of $170{\rm K}\lesssim T\lesssim250$ K in Appendix~\ref{sec:exhaustion}.

Before closing this section, we consider the $c$-dependence of $L_{\rm 12}$. In the extremely low-$T$ region where the impurity-band 
conduction dominates as seen in the inset of Fig.~\ref{fig:L12}, $|L_{\rm 12}|$ increases with $c$, in contrast to the high-$T$ $|L_{\rm 12}|$, 
which originates from the opposite tendency of the $c$-dependence of $\alpha_{\rm c}(E)$ for the conduction-band and impurity-band energy regions [see Figs.~\ref{fig:alpha}(b) and~\ref{fig:alpha}(c)].

\subsection{Seebeck Coefficient $S$~\label{sec:S}}

\begin{figure}[t]
 \begin{center}
\includegraphics[keepaspectratio=true,width=80mm]{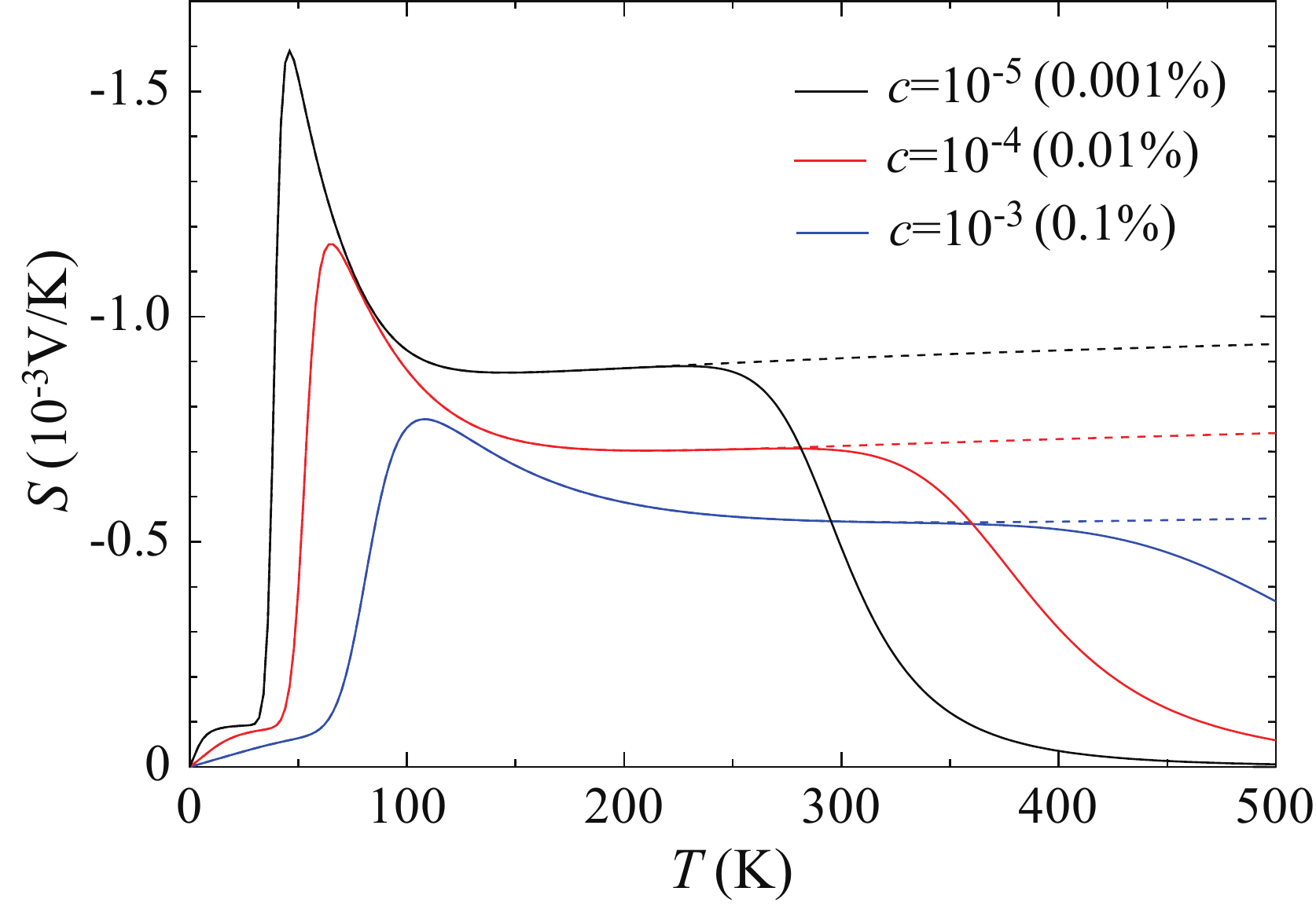}
  \end{center}
\caption{(Color online) Temperature dependence of the Seebeck coefficient $S$ of N-substituted (20,0) SWCNTs with $c=10^{-3}$ 
(blue solid curve), $10^{-4}$ (red solid curve) and $10^{-5}$ (black solid curve). The dashed curves are $S$ where the valence band 
is not taken into account in the calculation. Note that the vertical axis is negative.}
\label{fig:S}
\end{figure}

The Seebeck coefficient $S$ can be calculated using the relation of $S=\frac{1}{T}\frac{L_{\rm 12}}{L_{\rm 11}}$ in Eq.~(\ref{eq:S}). 
Figure~\ref{fig:S} shows the $T$-dependence of the $S$ value of N-substituted (20,0) SWCNTs 
for $c=10^{-3}$ (blue solid curve), $10^{-4}$ (red solid curve) and $10^{-5}$ (black solid curve).
The dashed curves are $S$ where the valence band is not taken into account in the calculation~\cite{rf:TY-HF01}.
Here, we explain $S$ with a focus on the case of $c=10^{-5}$ (black solid curve) as an example.
In the low-$T$ region, $|S|$ increases sharply near 30 K due to the rapid increase in $|L_{\rm 12}|$ near 30 K, as shown in the inset of Fig.~\ref{fig:L12}.
At extremely low $T$, much lower than 30 K, $|S|$ is proportional to $T$ in accordance with the Mott formula~\cite{rf:mott}, despite 
the impurity band conduction that cannot be described by the BTT~\cite{rf:TY-HF01}.
As $T$ increases, $|S|$ deviates rapidly upward from the Mott formula and has a large peak at $T\sim$50 K and then decreases with further increase in $T$.
The large peak originates from the thermal excitation from the impurity band to the conduction band with a small $m^*$, which is a different mechanism from a large $S$ of 1D semiconductors with pudding-mold-type band, $i. e,$ a large $m^*$~\cite{rf:pudding1,rf:pudding2}.
The $T$ dependence of $|S|$ can be explained in terms of the $T$ dependence of $L_{\rm 12}/L_{\rm 11}$.
As shown in the inset of Fig.~\ref{fig:L11}, $L_{\rm 11}$ begins to increase sharply at $T\sim$40 K. The $|S|$ peak at $T\sim$50 K indicates that 
$L_{\rm 12}/L_{\rm 11}$ is proportional to $T$, {\it i.e.},  $dS/dT=0$. Beyond $T\sim$50 K, the $T$-dependence of $L_{\rm 12}/L_{\rm 11}$ becomes 
weaker than $T$ linear and $|S|$ decreases with $T$.  In the region of $170{\rm K}\lesssim T\lesssim$250 K, which corresponds to the exhaustion region, 
$|S|$ is insensitive to $T$ because the conduction band electron density is almost constant. Over $T\sim$250 K entering the intrinsic region, 
$|S|$ decreases rapidly and approaches zero at the high-$T$ limit where the chemical potential $\mu$ is located at the center of the band gap, as 
represented in Fig.~\ref{fig:mu}. This is because $S$ due to conduction band electrons is perfectly cancelled out by valence-band 
holes in the limit $T\rightarrow\infty$. Similar features are evident in the red and blue solid curves.

\subsection{Power Factor $PF$~\label{sec:PF}}
\begin{figure}[t]
 \begin{center}
\includegraphics[keepaspectratio=true,width=80mm]{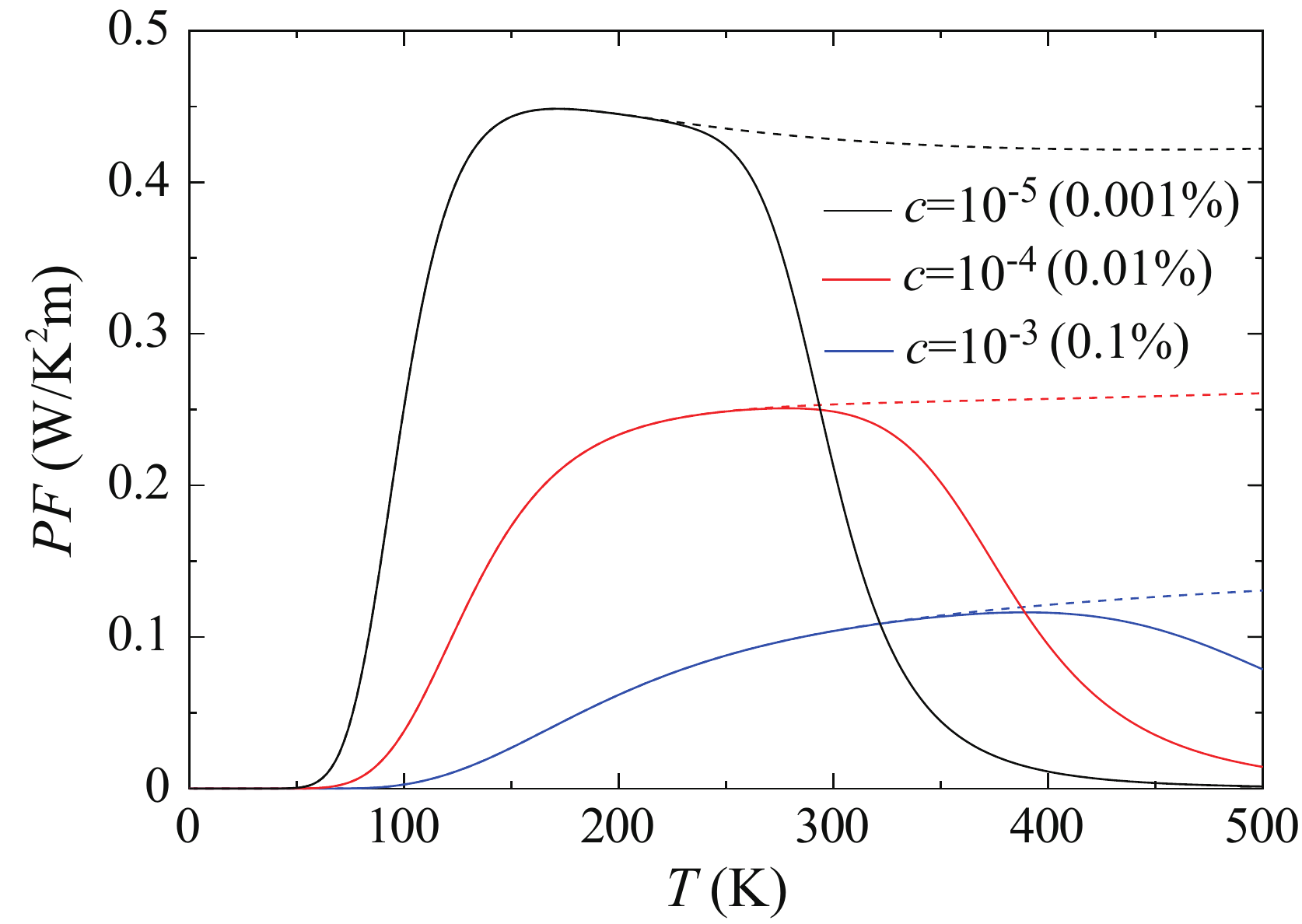}
  \end{center}
\caption{(Color online) Temperature dependence of the $PF$ for N-substituted (20,0) SWCNTs with $c=10^{-3}$ (blue solid curve), 
$10^{-4}$ (red solid curve) and $10^{-5}$ (black solid curve). The dashed curves show the $PF$s where the valence band is not taken into account in the calculation.
}
\label{fig:PF}
\end{figure}

The power factor ($PF$) can be calculated using the relation of $PF=L_{\rm 11}S^2$ in Eq.~(\ref{eq:PF}). Figure~\ref{fig:PF} shows the $T$-dependence of 
the $PF$ for N-substituted (20,0) SWCNTs with $c=10^{-3}$ (blue solid curve), $10^{-4}$ (red solid curve) and $10^{-5}$ (black solid curve).
Here, we explain the $PF$ with a focus on the case of $c=10^{-5}$ (black solid curve) as an example. Figure~\ref{fig:PF} shows that the $PF$ increases rapidly from 
$T\sim$30 K, at which $S$ rises sharply, as shown in Fig.~\ref{fig:S}. In the exhaustion region of $170{\rm K}\lesssim T\lesssim$250 K, the $PF$ shows weak 
dependence with respect to $T$ because the $T$-dependency of $L_{\rm 11}$ and $S$ are weak in this region. When $T$ exceeds approximately 250 K, entering 
the intrinsic region, the $PF$ drops rapidly and goes to zero due to the sharp decrease in $S$.
Similar $T$-dependence of the $PF$ can be observed for $c=10^{-4}$ (red solid curve) and $10^{-3}$ (blue solid curve), as represented in Fig.~\ref{fig:PF}.
In addition, the characteristic temperatures at which the solid curves deviate from the dashed curves shift toward high $T$ as $c$ increases.
Due to this shift, the optimal concentration $c_{\rm opt}$ that gives the maximum $PF$ is dependent on $T$.

\begin{figure}[t]
 \begin{center}
\includegraphics[keepaspectratio=true,width=80mm]{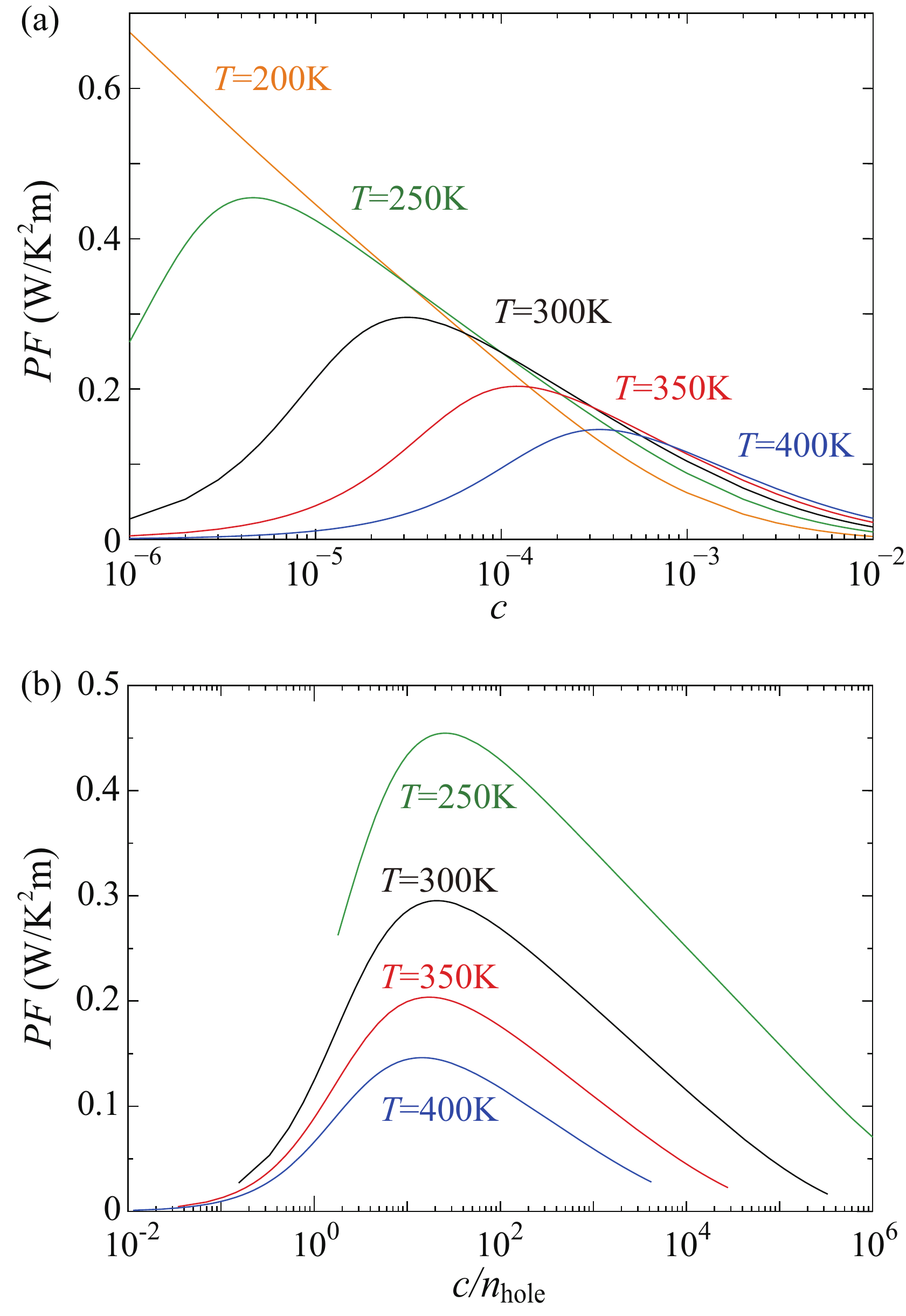}
  \end{center}
\caption{(Color online) The power factor $PF$ of N-substituted (20,0) SWCNTs at various temperatures as a function of the N concentration $c$ (a), and $c/n_{\rm hole}$ (b), where $n_{\rm hole}$ is the number of valence-band holes.
Each $PF$ curve has the maximum at $c_{\rm opt}/n_{\rm hole}\sim20$, such as $c_{\rm opt}/n_{\rm hole}$=26.5 at 250~K, 20.5 at 300~K, 16.7 at 350~K and 14.7 at 400~K.
}
\label{fig:PF_K}
\end{figure}

To show $c_{\rm opt}$ at a fixed temperature, we present the $c$-dependence of the $PF$ within $10^{-6}\leq c\leq10^{-2}$ at various temperatures in Fig.~\ref{fig:PF_K}(a).
At 200 K, the $PF$ increases monotonically with a decrease in $c$ within the present range of $c$.
This is because the thermal excitation from the valence band to the conduction one is negligible and
the monotonic increase of the $PF$ is given by $PF\propto(\ln c)^2$ ($c\ll1$) as discussed for N-substituted (10,0) SWCNTs in our previous report~\cite{rf:TY-HF01}.
In contrast, at $T=$250~K, 300 K, 350~K and 400 K, the $PF$s exhibit the maximum values at $c_{\rm opt}=4.7\times10^{-6}$, $3.1\times10^{-5}$, $1.2\times10^{-4}$ and $3.4\times10^{-4}$, respectively (see Table~\ref{t1}).
Thus, we can see that $c_{\rm opt}$ increases with increasing $T$.
In order to clarify what determines the value of $c_{\rm opt}$,
we show $PF$ as a function of $c/n_{\rm hole}$ in Fig.~\ref{fig:PF_K}(b).
Here, $n_{\rm hole}$ is the number of valence-band holes, which is defined by
\begin{eqnarray}
n_{\rm hole}=4\int_{-\infty}^{E_{\rm v}}\rho(E)(1-f(E))dE,
\end{eqnarray}
where $E_{\rm v}(<0)$ is the valence-band top and the factor 4 comes from the spin degeneracy and the orbital degeneracy.
As shown in Fig.~\ref{fig:PF_K}(b), each $PF$ curve exhibits a peak at $c_{\rm opt}/n_{\rm hole}\sim20$, which means that $PF$ becomes the maximum when the N concentration reaches about 20 times the number of thermally excited holes.
Note that this condition ($c_{\rm opt}/n_{\rm hole}\sim20$) is not satisfied for N-substituted (10,0) SWCNTs with $c=10^{-2}\sim10^{-5}$ at $T\leq 400$~K.
As a result, $PF$ does not show the maximum for the condition of $c=10^{-2}\sim10^{-5}$ and $T\leq 400$~K as shown in Appendix~\ref{sec:previous}.

\begin{table}
\caption{List of maximum power factor $PF_{\rm max}$ and $c_{\rm opt}$ for various temperature.}
\label{t1}
\begin{center}
\begin{tabular}{c||c|c|c|c}
&250~K&300~K&350~K&400~K\\
\hline\hline
$PF_{\rm max}$ ($\rm W/K^2m$)&0.45&0.30&0.20&0.15\\
\hline
$c_{\rm opt}$&$4.7\times10^{-6}$&$3.1\times10^{-5}$&$1.2\times10^{-4}$&$3.4\times10^{-4}$\\
\hline
\end{tabular}
\end{center}
\end{table}

\begin{figure}[t]
  \begin{center}
  \includegraphics[keepaspectratio=true,width=80mm]{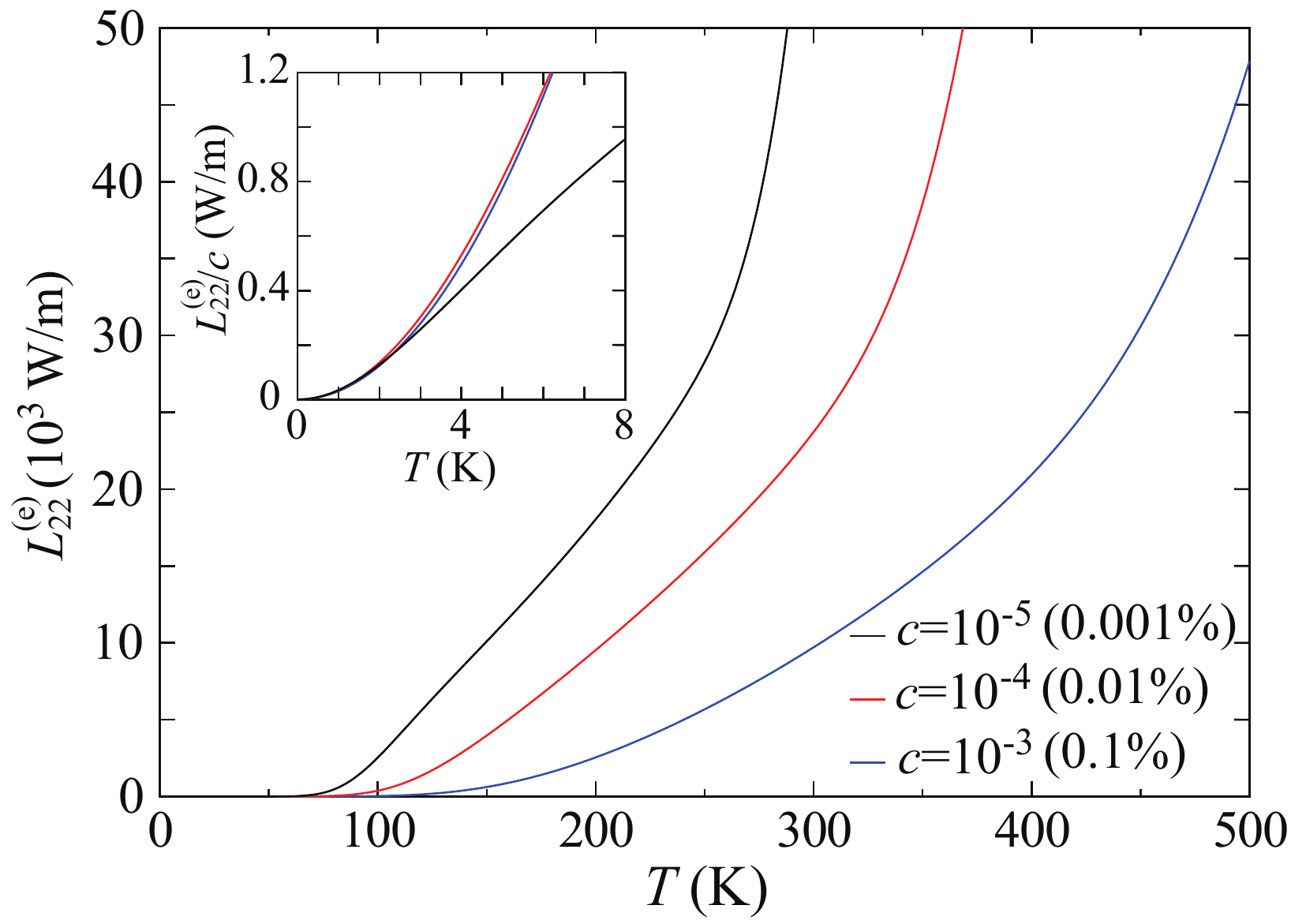}
  \end{center}
\caption{(Color online) Temperature dependence of electronic thermal conductivity $L_{22}$ of N-substituted (20,0) SWCNTs 
with $c=10^{-3}$ (blue solid curve), $10^{-4}$ (red solid curve) and $10^{-5}$ (black solid curve). The inset is the enlarged view of low-$T$ region.}
\label{fig:L22}
\end{figure}

\subsection{Electronic Thermal Conductivity $\lambda_{\rm e}$~\label{sec:L22}}
Thermal conductivity is known to be due to electrons and phonons: the former, electronic thermal conductivity, $\lambda_{\rm e}$,
is defined as follows,
\begin{eqnarray}
\lambda_{\rm e}=\frac{1}{T}\left(L_{22}^{\rm (e)}-\frac{L_{12}L_{21}}{L_{11}}\right),
\label{eq:lambda}
\end{eqnarray}
where $L_{22}^{\rm (e)}$ is given by 
\begin{eqnarray}
L_{22}^{\rm (e)}&=&\frac{1}{e^2}\int_{-\infty}^\infty \!\!\!dE\left(-\frac{\partial f(E-\mu)}{\partial E}\right)(E-\mu)^2\alpha(E)
\label{eq:L22}
\end{eqnarray}
in the present case of N-substituted SWCNTs (See Appendix~\ref{sec:derivation_L22}).

Figure~\ref{fig:L22} shows the $T$-dependence of $L_{22}^{\rm (e)}$ for N-substituted (20,0) SWCNTs with $c=10^{-3}$ (blue solid curve), 
$10^{-4}$ (red solid curve) and $10^{-5}$ (black solid curve). 
As seen in Fig.~\ref{fig:L22}, $L_{22}^{\rm (e)}$ increases monotonically with increasing $T$ for all $c$.
At a fixed $T$, $L_{22}^{\rm (e)}$ increases as $c$ decreases except for extremely low $T$ at which the impurity-band conduction is dominant.
This is due to the fact that $\alpha(E)$ in the conduction band contributing at finite $T$ because of thermal excitations is proportional to $1/c$ as shown in Fig. 5 (b).
On the other hand, $L_{22}^{\rm (e)}/c$ is proportional to $T^2$ and is independent of $c$ in the limit of low $T$ as seen
by the Sommerfeld expansion as
\begin{eqnarray}
L_{22}^{\rm (e)}\sim\frac{\pi^2k_B^2}{3e^2}\alpha(E_F)T^2,
\label{eq:sL22}
\end{eqnarray}
with $\alpha(E_{\rm F})\propto c$ at the Fermi energy $E_F$ lying in the impurity band (See Fig.~\ref{fig:alpha}(c)).

\begin{figure}[t]
  \begin{center}
  \includegraphics[keepaspectratio=true,width=80mm]{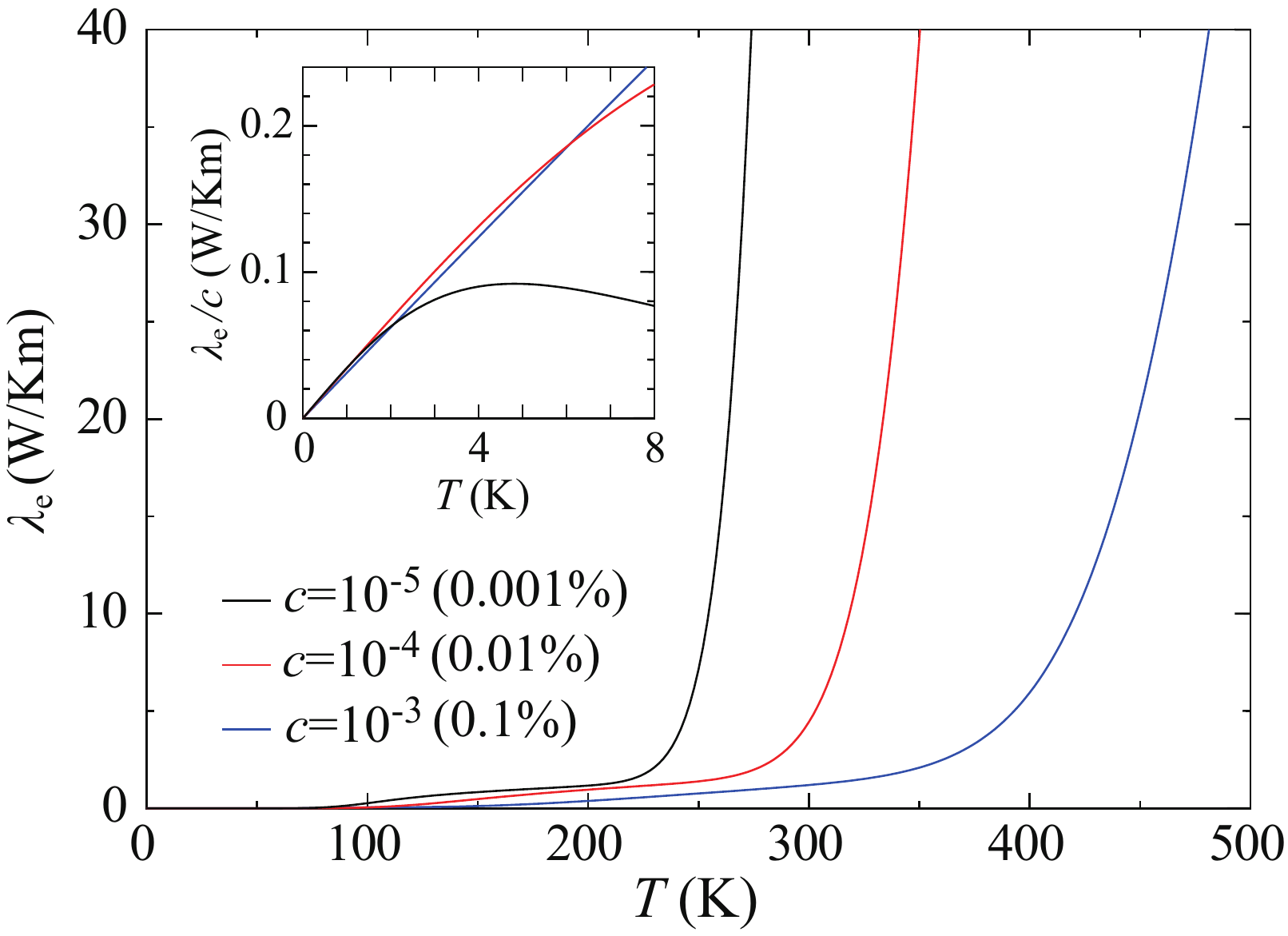}
  \end{center}
\caption{(Color online) Temperature dependence of electronic thermal conductivity $\lambda_{\rm e}$ of N-substituted (20,0) SWCNTs 
with $c=10^{-3}$ (blue solid curve), $10^{-4}$ (red solid curve) and $10^{-5}$ (black solid curve). The inset is the enlarged view of low-$T$ region.}
\label{fig:lambda}
\end{figure}

Figure \ref{fig:lambda} displays the $T$-dependence of  $\lambda_{\rm e}$ for $c=10^{-3}$ (blue solid curve), $10^{-4}$ (red solid curve) and 
$10^{-5}$ (black solid curve). Similar to $L^{\rm (e)}_{22}$, $\lambda_{\rm e}$ increases as $T$ increases and as $c$ decreases except 
for extremely low $T$, while at an extremely low $T$, $\lambda_{\rm e}/c$ is proportional to $T$ and 
is independent of $c$ as shown in the inset of Fig.~\ref{fig:lambda}. This can be understood by the Sommerfeld expansion as
\begin{eqnarray}
\lambda_{\rm e}\approx\frac{\pi^2k_B^2}{3e^2}\alpha(E_F)T,
\label{eq:slambda}
\end{eqnarray}
and $\alpha(E_{\rm F})\propto c$ as shown in Fig.~\ref{fig:alpha}(c). 
As seen from Fig. \ref{fig:PF}, the contribution of second term $L_{12}L_{21}/(TL_{11})=PF\times T$ in Eq.~(\ref{eq:lambda}) is negligible in 
comparison to the first term $L_{22}^{\rm (e)}/T$ except for the exhaustion region.

\begin{figure}[t]
  \begin{center}
  \includegraphics[keepaspectratio=true,width=80mm]{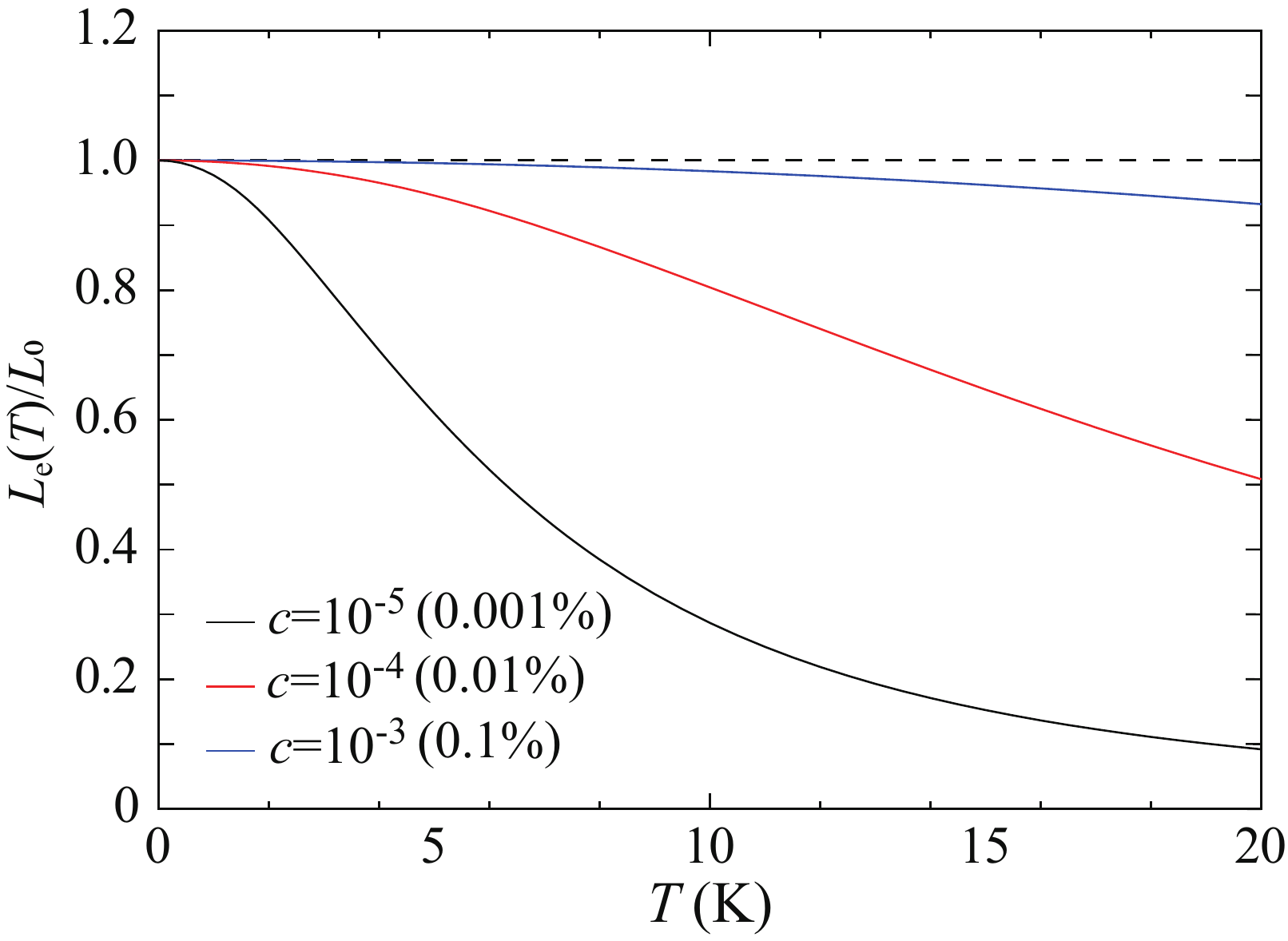}
  \end{center}
\caption{(Color online) Temperature dependence of Lorenz ratio $L_{\rm e}(T)$ scaled by the universal Lorenz number $L_0\equiv\pi^2k_B^2/(3e^2)$ of 
N-substituted (20,0) SWCNTs with $c=10^{-3}$ (blue solid curve), $10^{-4}$ (red solid curve) and $10^{-5}$ (black solid curve).
The dashed line represents the Lorentz number $L_{\rm e}(T)=L_0$.}
\label{fig:L}
\end{figure}

Figure~\ref{fig:L} illustrates the low-$T$ behavior of electronic contribution to the Lorenz ratio $L_{\rm e}(T)\equiv\lambda_{\rm e}(T)/(TL_{11}(T))$ 
scaled by the universal Lorenz number $L_0\equiv\pi^2k_B^2/(3e^2)$ for $c=10^{-3}$ (blue solid curve), $10^{-4}$ (red solid curve) and $10^{-5}$ 
(black solid curve). All curves in Fig.~\ref{fig:L} approach unity  in the low-$T$ limit.
This means that the Wiedemann-Franz law holds even for the impurity-band conduction.
As $T$ increases, $L_{\rm e}(T)$ deviates downward from $L_0$ in proportion to $T^2$ as
\begin{eqnarray}
\frac{L_{\rm e}(T)}{L_0}\approx1-\frac{\pi^2}{3}\left\{\frac{\alpha''(E_F)}{2\alpha(E_F)}+\left(\frac{\alpha'(E_F)}{\alpha(E_F)}\right)^2\right\}(k_BT)^2.
\label{eq:slambda}
\end{eqnarray}

From Fig.~\ref{fig:lambda}, we note $\lambda_{\rm e}$ is much smaller than the phonon thermal conductivity, $\lambda_{\rm ph}$.
It is known that room temperature $\lambda_{\rm ph}$ of SWCNTs without N impurities is of the order of 1,000~W/Km~\cite{rf:lambda1,rf:lambda2,rf:lambda3,rf:lambda4,rf:lambda5,rf:lam_cnt1,rf:lam_cnt2,rf:lam_cnt3},
which is comparable to that of N-substituted SWCNT with dilute N concentration~\cite{rf:sasa1,rf:zhu}.
Hence in the figure of merit $ZT=(PF/\lambda)T$, $PF$ is determined by electrons while $\lambda$ by phonons, $ZT\approx(PF/\lambda_{\rm ph})T$,
and then the analysis of optimal condition for $PF$ applies also for $ZT$ resulting in $ZT\sim 0.1$ for N-substituted (20,0) SWCNTs with
$c_{\rm opt}=3.1\times 10^{-5}$ at 300K.

\section{Summary~\label{sec:4}}
The thermoelectric effects of N-substituted SWCNTs were investigated using the Kubo-L{\" u}ttinger theory combined with the Green's function technique.
We have clarified the temperature dependence of the electrical conductivity $L_{\rm 11}$ and thermoelectrical conductivity $L_{\rm 12}$, as well as 
the Seebeck coefficient $S$ and power factor $PF$ for a wide temperature range from the ionization region to the intrinsic region through 
the exhaustion region. $S$ and $PF$ decrease rapidly toward zero around a crossover temperature from the exhaustion region to the intrinsic region, 
and the crossover temperature shifts toward higher temperature with an increase in the impurity concentration.
Due to this doping dependence of the shift of the crossover temperature, the optimal impurity concentration $c_{\rm opt}$ that gives 
the maximum $PF$ changes depending on temperature.
As shown in Table~\ref{t1}, we have determined $c_{\rm opt}$ for various temperature for N-substituted (20,0) SWCNTs.
In addition, using the Sommerfeld-Bethe expression of $L_{22}^{\rm (e)}$, we elucidate the temperature dependence of 
$\lambda_{\rm e}\equiv(L_{22}^{\rm (e)}-L_{12}L_{21}/L_{11})/T$ and show that the Wiedemann-Franz law for $\lambda_{\rm e}/L_{11}$ 
is valid in the limit of low $T$ even for the impurity-band conduction.
The optimal condition for $PF$ applies also for the figure of merit $ZT$ because the electronic thermal conductivity $\lambda_{\rm e}$ is much smaller than the phonon thermal conductivity $\lambda_{\rm ph}$.
We estimate $ZT\sim 0.1$ for N-substituted (20,0) SWCNTs with $c_{\rm opt}=3.1\times 10^{-5}$ and $\lambda_{\rm ph}=1,000$W/Km at room temperature.

Finally, we note that the results obtained in the present study can also be applied to boron-substituted SWCNTs by replacement of the impurity potential from an attractive potential to a repulsive potential.

\section*{Acknowledgments}
\begin{acknowledgment}
This work was supported by JSPS KAKENHI through Grant Numbers JP18H01816 and JP20K15117 and 
by The Thermal and Electric Energy Technology Foundation (TEET).

\end{acknowledgment}

\appendix
\section{Nitrogen Concentration Dependence of Power Factor of N-substituted (10,0) SWCNTs~\label{sec:previous}}
\begin{figure}[t]
 \begin{center}
\includegraphics[keepaspectratio=true,width=80mm]{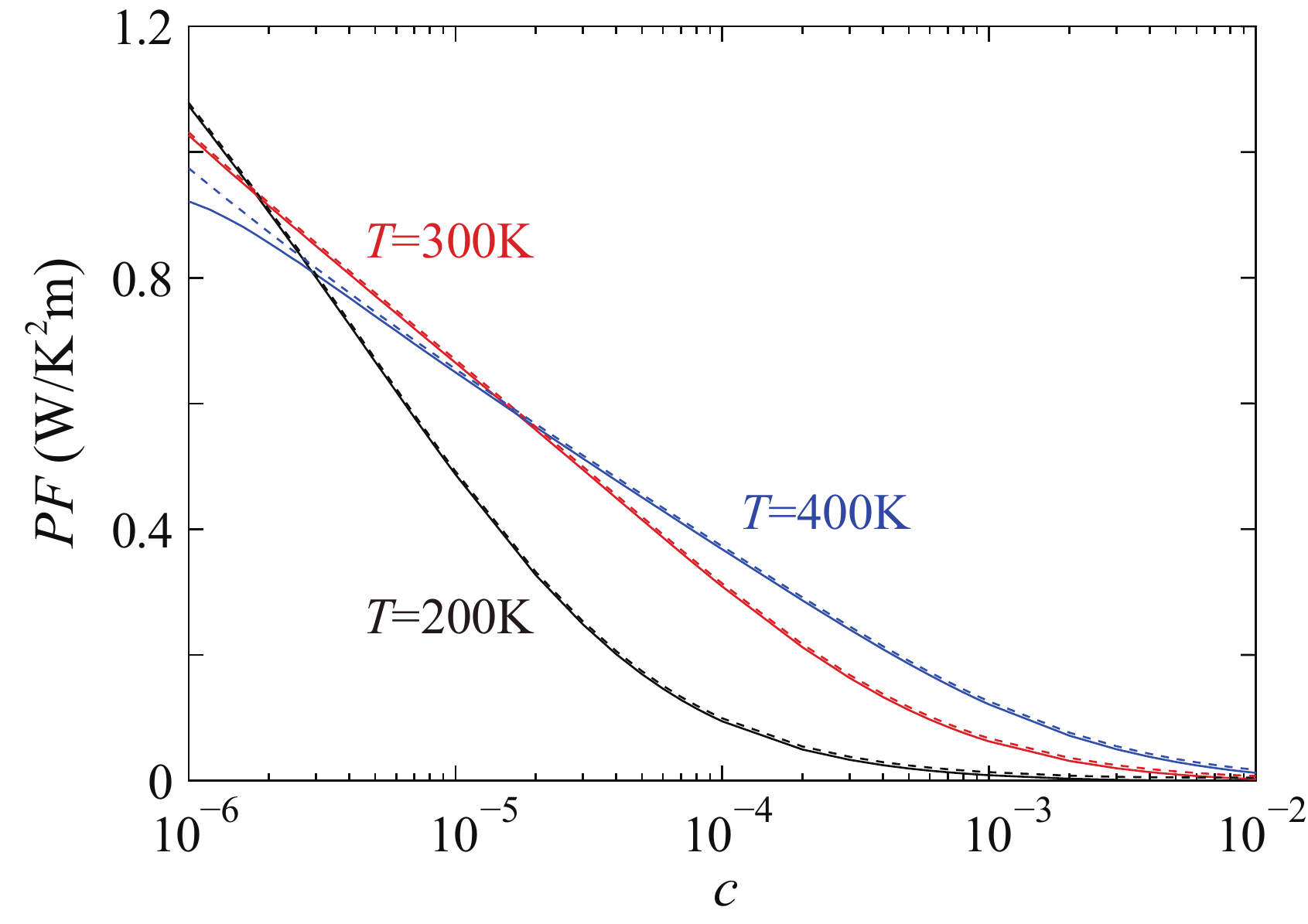}
  \end{center}
\caption{(Color online)
N concentration dependence of the $PF$ for N-substituted (10,0) SWCNTs at $T=200$~K (black curve), 300~K (red curve) and 400~K (blue curve).
The dashed curves show $PF$s where the valence band is not taken into account in the calculation.}
\label{fig:previous}
\end{figure}

In this Appendix, we discuss the contribution of thermal excitation from the valence band to the conduction band to the $PF$ of an N-substituted (10,0) SWCNT.
Figure~\ref{fig:previous} shows the $c$-dependence of $PF$ of the N-substituted (10,0) SWCNT at $T=200$~K, 300~K and 400~K.
The solid curves indicate $PF$s for systems including N-impurity bands with both conduction and valence bands self-consistently as discussed in Sec.~\ref{sec:2.3} and the dashed curves are $PF$s shown in Fig.8(c) of the previous paper~\cite{rf:TY-HF01} where the valence band is not incorporated into electronic states of N-substituted SWCNTs.
As seen in Fig.~\ref{fig:previous}, the solid curves fit the dashed curves $c\gtrsim 10^{-5}$ even at $T=400$~K.
This means that the valence band does not contribute to the $PF$ of N-substituted (10,0) SWCNT.
In addition, all the solid curves increase with decreasing $c$ within $c\geq 10^{-6}$ and do not exhibit the maximum, which is different from the N-substituted (20,0) SWCNT discussed in this paper.

\section{Temperature Dependence of Electron Number in Conduction Band~\label{sec:n_c}}
\begin{figure}[t]
 \begin{center}
\includegraphics[keepaspectratio=true,width=80mm]{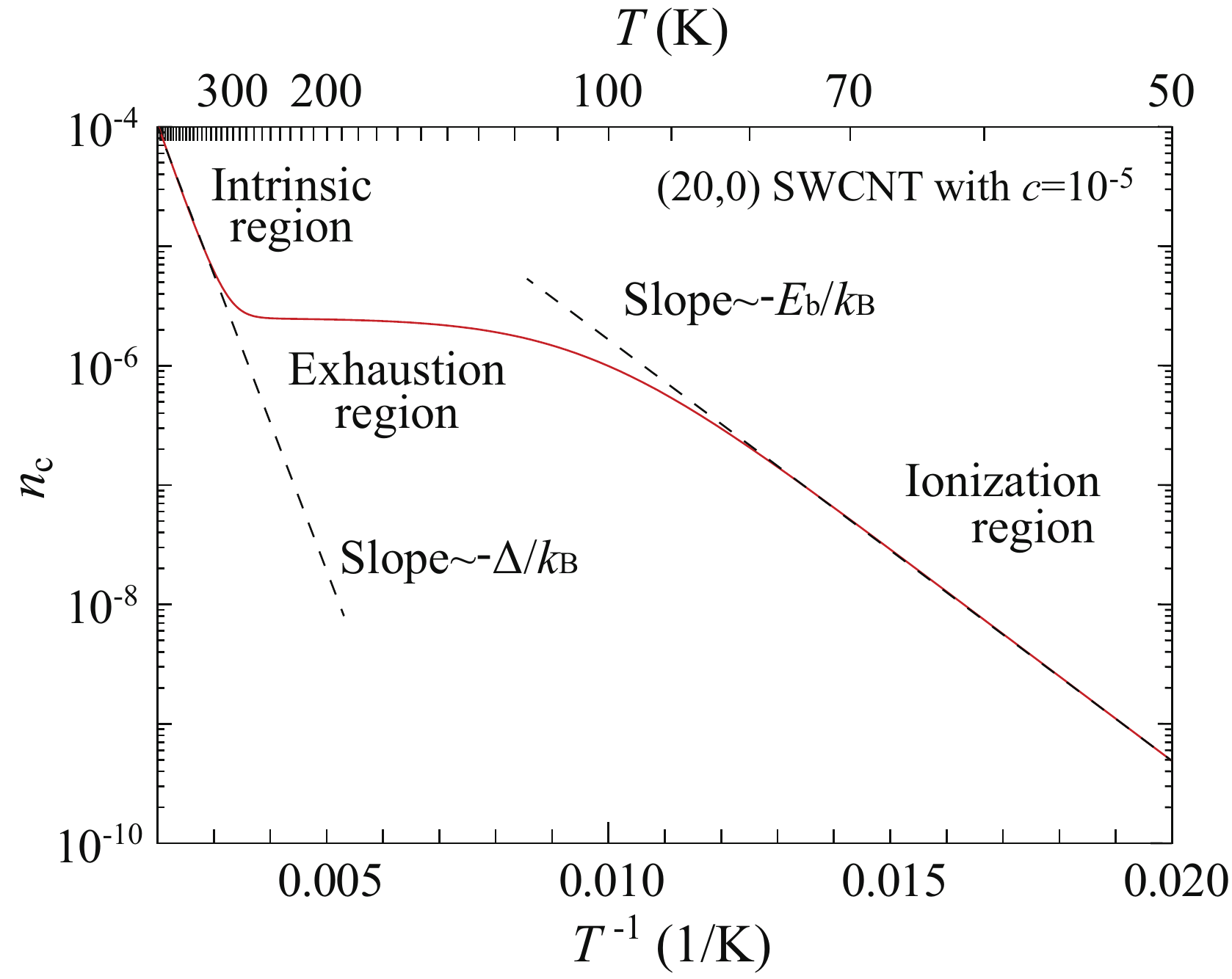}
  \end{center}
\caption{(Color online)
Temperature dependence of the electron number $n_{\rm c}$ per unit cell for each spin and each orbital in the conduction band of N-substituted 
(20,0) SWCNTs with $c=10^{-5}$.
}
\label{fig:n_c}
\end{figure}

Figure~\ref{fig:n_c} shows the $T$-dependence of the conduction band electron number $n_{\rm c}$ per unit cell for each spin and each orbital of 
the N-substituted (20,0) SWCNT with $c=10^{-5}$. 
Here, $n_{\rm c}$ is determined by
\begin{eqnarray}
n_{\rm c}=\int_{E_{\rm c}}^{\infty}\rho(E)f(E)dE,
\end{eqnarray}
where $E_{\rm c}$ is the conduction-band bottom, which can be determined by the condition $dx/d\sigma_{\rm c}=0$.
The $T$-dependence of $n_{\rm c}$ has three regions: the ionization region at low $T$,
the exhaustion region at middle $T$ and the intrinsic region at high $T$. In the ionization region, most N atoms (donors) still capture valence electrons, 
{\it i.e.}, they are not thermally ionized, and the $T$-dependence of $n_{\rm c}$ is given by $n_{\rm c}\sim \exp(-E_{\rm b}/k_{\rm B}T)$ (see Fig.~\ref{fig:n_c}).
In the exhaustion region, most N atoms are thermally ionized, and the valence band electrons are still frozen out. In this case, $n_{\rm c}$ is almost equal to 
the density of N atoms and is independent of $T$ (see Fig.~\ref{fig:n_c}). In the intrinsic region, the valence band electrons are thermally excited from 
the valence band to the conduction band, and the $T$-dependence of $n_{\rm c}$ is given by $n_{\rm c}\sim \exp(-\Delta/k_{\rm B}T)$ (see Fig.~\ref{fig:n_c}).

\section{Spectral Conductivity in Conduction- and Valence-Band Energy Regions~\label{sec:t_mat}}
We discuss the spectral conductivity of N-substituted SWCNTs with low $c$ in the conduction- and valence-band energy regions.
In the case of low $c$, the self-energy $\Sigma_{\rm c}(E)$ for the conduction-band electrons ($E>\Delta$) in Eq.~(\ref{eq:self-consistent_Sigma}) can be described by
\begin{eqnarray}
\Sigma_{\rm c}(E)&=&\frac{cV_{\rm 0}}{1+i\frac{V_{\rm 0}/2}{\sqrt{t(E-\Delta)}}}\\
&\equiv&A_{\rm c}(E)-i\frac{\hbar}{2\tau_{\rm c}(E)}
\label{eq:alpha_non}
\end{eqnarray}
within the $t$-matrix approximation (self-consistency is not necessary because of low $c$), 
where the energy shift $A_{\rm c}(E)$ and the relaxation time $\tau_{\rm c}(E)$ are respectively defined by
\begin{eqnarray}
A_{\rm c}(E)=\frac{-c|V_{\rm 0}|}{1+\frac{(V_0/2)^2}{t(E-\Delta)}}
\label{eq:A}
\end{eqnarray}
and
\begin{eqnarray}
\tau_{\rm c}(E)=\frac{\hbar}{cV_{\rm 0}^2}\left(\sqrt{t(E-\Delta)}+\frac{(V_0/2)^2}{\sqrt{t(E-\Delta)}}\right).
\label{eq:tau_c}
\end{eqnarray}
Using Eqs.~(\ref{eq:sc_cv}), (\ref{eq:retarded_G}) and (\ref{eq:alpha_non}), $\alpha_{\rm c}(E)$ can be written as the BTT expression of
\begin{eqnarray}
\alpha_{\rm c}(E)= 4\frac{e^2}{V}N_{\rm unit}v_{\rm c}^2(E)\rho_{\rm c}(E)\tau_{\rm c}(E)
\label{eq:alpha_BTT}
\end{eqnarray}
where the group velocity $v_{\rm c}$ of a conduction-band electron is
\begin{eqnarray}
v_{\rm c}(E)=\pm\sqrt{\frac{2(E-\Delta+|A_{\rm c}(E)|)}{m^*}}
\label{eq:velocity_ccc}
\end{eqnarray}
and the DOS per unit cell for each spin and each orbital of conduction-band electrons is
\begin{eqnarray}
\rho_{\rm c}(E)=\frac{a_z}{\pi\hbar}\sqrt{\frac{m^*}{2(E-\Delta+|A_{\rm c}(E)|)}}.
\label{eq:v_rho}
\end{eqnarray}
Substituting Eqs.~(\ref{eq:velocity_ccc}) and (\ref{eq:v_rho}) into Eq.~(\ref{eq:alpha_BTT}), we obtain
\begin{eqnarray}
\alpha_{\rm c}(E)=4\frac{e^2}{A\pi\hbar}\sqrt{\frac{2(E-\Delta+|A_{\rm c}(E)|)}{m^*}}\tau_{\rm c}(E).
\label{eq:alpha2}
\end{eqnarray}
In the low-$c$ case obeying $c\ll |V_0|/4t$ ({\it i.e.}, $E-\Delta\gg |A_{\rm c}(E)|$), the energy shift $A_{\rm c}(E)$ is negligible and thus
$\alpha_{\rm c}(E)$ can be written as
\begin{eqnarray}
\alpha_{\rm c}(E)=4\frac{e^2}{A\pi\hbar}\sqrt{\frac{2(E-\Delta)}{m^*}}\tau_{\rm c}(E).
\label{eq:alpha3}
\end{eqnarray}
Because $\tau_{\rm c}$ is inversely proportional to $c$ as seen in Eq.~(\ref{eq:tau_c}), $\alpha_{\rm c}(E)$ is also
inversely proportional to $c$, that is $\alpha_{\rm c}(E)\propto 1/c$.
Similarly, the spectral conductivity $\alpha_{\rm v}(E)$ for valence-band electrons ($E<-\Delta$) also obeys $\alpha_{\rm v}(E)\propto 1/c$.
Note that in the case of (20,0) SWCNT,  $|V_0|/4t=7.5\times 10^{-2}$ is much larger than $c\le 10^{-3}$ used in the present study.

\section{Temperature Dependence of $L_{\rm 11}$ and $L_{\rm 12}$ in the Exhaustion Region~\label{sec:exhaustion}}
According to our previous report~\cite{rf:TY-HF01}, the $T$-dependence of $L_{\rm 11}$ and $L_{\rm 12}$ for N-substituted SWCNTs in the exhaustion region are 
expressed as
\begin{eqnarray}
L_{\rm 11}=\frac{e^2a_z}{\hbar A}\sqrt{\frac{t}{\pi k_{\rm B}T}}\left(1+\frac{k_{\rm B}T}{E_{\rm b}}\right)
\label{eq:appL11}
\end{eqnarray}
and
\begin{eqnarray}
L_{\rm 12}=-\frac{ea_z}{\hbar A}\sqrt{\frac{tk_{\rm B}T}{\pi}}\left(1+2\frac{k_{\rm B}T}{E_{\rm b}}\right)+\frac{\mu}{e}L_{\rm 11}
\label{eq:appL12}
\end{eqnarray}
with $\mu=k_BT{\rm ln}(\frac{c}{2}\sqrt{\frac{\pi t}{k_{\rm B}T}})$.
We have confirmed that these analytical results are in good agreement with the numerical results for $L_{\rm 11}$ and $L_{\rm 12}$ based on 
the Kubo-L{\" u}ttinger theory shown in Figs.~\ref{fig:L11} and ~\ref{fig:L12}.

\section{Electronic Thermal Conductivity~\label{sec:derivation_L22}}
Under the electric field $\mathcal{E}$ and the temperature gradient $dT/dz$, the thermal current density $J_{\rm Q}$ is expressed as
\begin{eqnarray}
J_{\rm Q}=L_{21}\mathcal{E}-\frac{L_{22}}{T}\frac{dT}{dz},
\label{eq:JQ}
\end{eqnarray}
within the linear response with $\mathcal{E}$ and $dT/dz$. Here, $L_{21}$ is called $electrothermal$ $conductivity$ and is connected to 
$L_{12}$ by Onsager's reciprocal relation $L_{21}=L_{12}$~\cite{rf:Onsager}. $L_{22}$ includes the contributions from electrons and phonons,
that is $L_{22}^{\rm (e)}+L_{22}^{\rm (ph)}$. Thus, electronic thermal conductivity $\lambda_{\rm e}$ is given by
\begin{eqnarray}
\lambda_{\rm e}=\frac{1}{T}\left(L_{22}^{\rm (e)}-\frac{L_{12}L_{21}}{L_{11}}\right).
\label{eq:lambda2}
\end{eqnarray}
under the condition of $J=0$ ($i.e.,$ $\mathcal{E}=\frac{L_{12}}{TL_{11}}\frac{dT}{dz}$ from Eq.~(\ref{eq:J})).
According to Kubo's linear response theory, $L_{22}^{\rm (e)}$ can be obtained as
\begin{eqnarray}
L_{22}^{\rm (e)}=-\displaystyle \lim_{\omega\to0}\frac{\chi_{22}^{\rm R}(\omega)-\chi_{22}^{\rm R}(0)}{i\omega},\\
\label{eq:L22_2}
\chi_{22}^{\rm R}(\omega)=\left.\chi_{22}(i\omega_{\lambda})\right|_{i\omega_{\lambda}\rightarrow \hbar\omega+i\delta},
\label{eq:chi}
\end{eqnarray}
where $\chi_{22}(i\omega_\lambda)$ is the $J_{\rm Q}$-$J_{\rm Q}$ correlation function, expressed as
\begin{eqnarray}
\chi_{22}(i\omega_{\lambda})=\frac{1}{V}\int_0^\beta d\tau\langle T_\tau\{J_{\rm Q}(\tau)J_{\rm Q}(0)\}\rangle e^{i\omega_\lambda\tau},
\label{eq:chi2}
\end{eqnarray}
where $\beta\equiv1/(k_BT)$ is the inverse temperature, $T_\tau$ is the imaginary-time-ordering operator, $\langle\cdots\rangle$ denotes 
the thermal average in equilibrium, and $V$ is the volume of a system with $J_{\rm Q}$ being the thermal current density.
In the present case, where electrons are scattered by elastic impurities, $J_{\rm Q}$ is given by
\begin{eqnarray}
J_{\rm Q}(\tau)=\sum_{k}\left[\bar\Phi_k(\tau)
   \left(\begin{array}{cc}
      (\epsilon^{(+)}_k-\mu)v^{(+)}_k & 0 \\
      0 & (\epsilon^{(-)}_k-\mu)v^{(-)}_k  \\
    \end{array}\right)
  \Phi_k(\tau)\right.\nonumber\\
\left.+\frac{1}{2}\sum_q
\bar\Phi_{k+q}(\tau)
   \left(\begin{array}{cc}
      u(q)(v^{(+)}_k+v^{(+)}_{k+q}) & 0 \\
      0 & u(q)(v^{(-)}_k+v^{(-)}_{k+q})  \\
    \end{array}\right)
  \Phi_k(\tau)\right]
\label{eq:JQ2}
\end{eqnarray}
where $v_k^{(\pm)}=\pm\hbar k/m^*$, $u(q)=V_0\sum_{\langle j\rangle} e^{-iqj}/N$,
\begin{eqnarray}
\bar\Phi_{k}(\tau)=\left(\begin{array}{cc}
e^{\tau\mathcal{H}}c^\dagger_ke^{-\tau\mathcal{H}},& e^{\tau\mathcal{H}}d^\dagger_ke^{-\tau\mathcal{H}}\\
    \end{array}\right)
    \end{eqnarray}
    and
    \begin{eqnarray}
\Phi_{k}(\tau)=\left(\begin{array}{c}
      e^{\tau\mathcal{H}}c_ke^{-\tau\mathcal{H}} \\
      e^{\tau\mathcal{H}}d_ke^{-\tau\mathcal{H}} \\
    \end{array}\right).
\end{eqnarray}
Substituting Eq.~(\ref{eq:JQ2}) into Eq.~(\ref{eq:chi2}) and performing a similar procedure as Ref.\citen{rf:Jonson_1980} by Jonson and Mahan,
we can straightforwardly obtain the SB type expression of $L_{22}^{\rm(e)}$ in Eq.~(\ref{eq:L22}).

\end{document}